\documentclass[12pt]{amsart}
\usepackage{amsmath}
\usepackage{amsxtra}
\usepackage{amscd}
\usepackage{amsthm}
\usepackage{amsfonts}
\usepackage{amssymb}
\usepackage{eucal}
\usepackage{epsfig}
\usepackage{graphics}
\def\({\left(}
\def\){\right)}
\newcommand{\Rdr}{R_\mathrm{dress}}

\newcommand{\phib}{\mbox{\boldmath$\phi$}}
\newcommand{\betab}{\mbox{\boldmath$\beta$}}
\newcommand{\gammab}{\mbox{\boldmath$\gamma$}}

\newcommand{\rhob}{\rho^\mathrm{sc}}
\newcommand{\omegab}{\omega^\mathrm{sc}}
\newcommand{\Tb}{T^\mathrm{sc}}
\newcommand{\Qb}{Q^\mathrm{sc}}

\newcommand{\taub}{\mbox{\boldmath$\tau$}}

\newcommand{\cb}{\mathbf{c}}
\newcommand{\bb}{\mathbf{b}}

\newcommand{\tb}{\mathbf{t}}

\newcommand{\nn}{\nonumber}
\newcommand{\bea}{\begin{eqnarray}}
\newcommand{\ena}{\end{eqnarray}}
\def\bel{\begin{eqnarray}}
\def\enl{\end{eqnarray}}
\newcommand{\be}{\begin{eqnarray*}}
\newcommand{\en}{\end{eqnarray*}}
\newcommand{\ba}{\begin{array}}
\newcommand{\ea}{\end{array}}

\newcommand{\R}{{\mathbb R}}

\newcommand{\slth}{\widehat{\mathfrak{sl}}_2}
\newcommand{\res}{{\rm res}}

\newcommand{\Tr}{{\rm Tr}}

\renewcommand{\Im}{\mathop{\rm Im}}
\newenvironment{tenumerate}{
  \begin{enumerate}
  
  }{\end{enumerate}}
\newcommand{\bi}{\begin{tenumerate}}
\newcommand{\ei}{\end{tenumerate}}
\newcommand{\isoto}[1][]%
{{\mathop{\buildrel{\sim}\over\longrightarrow}\limits_{#1}}}

\def\[{\left[}
\def\]{\right]}
\newcommand{\la}{\lambda}

\newcommand{\al}{\alpha}

\newcommand{\z}{\zeta}

\numberwithin{equation}{section}
\newtheorem{thm}{Theorem}[section]

\newtheorem{rem}[thm]{Remark}

\newcommand{\cyl}{\mbox{$C\hskip-1pt yl$}}

\def\half{\textstyle{\frac  1 2}}

\def\bi{\mathbf{i}}

\begin{document}

\begin{title}[Grassmann Structure in XXZ Model]
{Hidden Grassmann Structure in the XXZ Model IV:
CFT limit
}
\end{title}
\date{\today}
\author{H.~Boos, M.~Jimbo, T.~Miwa and  F.~Smirnov}
\address{HB: Physics Department, University of Wuppertal, D-42097,
Wuppertal, Germany}\email{boos@physik.uni-wuppertal.de}
\address{MJ: Department of Mathematics, 
Rikkyo University, Toshima-ku, 
Tokyo 171-8501, Japan}
\email{jimbomm@rikkyo.ac.jp}
\address{TM: Department of 
Mathematics, Graduate School of Science,
Kyoto University, Kyoto 606-8502, 
Japan}\email{tmiwa@kje.biglobe.ne.jp}
\address{FS\footnote{Membre du CNRS}: Laboratoire de Physique Th{\'e}orique et
Hautes Energies, Universit{\'e} Pierre et Marie Curie,
Tour 16 1$^{\rm er}$ {\'e}tage, 4 Place Jussieu
75252 Paris Cedex 05, France}\email{smirnov@lpthe.jussieu.fr}

\begin{abstract}

The Grassmann structure of the critical XXZ spin chain
is studied in the limit to conformal field theory. 
A new description of Virasoro Verma modules is proposed 
in terms of Zamolodchikov's integrals of motion and 
two families of fermionic creation operators. 
The exact relation to the usual Virasoro description is found up to level 6.
\end{abstract}

\maketitle

\section{Introduction}\label{intro}

In the present paper we continue the 
series of works \cite{HGSI,HGSII,HGSIII} 
on the XXZ model.
In \cite{HGSIII} we considered it 
in the presence of the Matsubara direction,
or equivalently the six vertex model on a cylinder. 
We computed the normalised partition function 
with a defect localised between two horizontal lines, 
which corresponds to an insertion of a quasi-local operator:
\begin{align}
Z^{\kappa}
\Bigl\{q^{2\al S(0)}\mathcal{O}
\Bigr\}=
\frac{\Tr _{\mathrm{S}}
\Tr _{\mathbf{M}}\Bigl(T_{\mathrm{S},\mathbf{M}}q^{2\kappa S+2\al S(0)}\mathcal{O}\Bigr)}
{\Tr _{\mathrm{S}}\Tr _\mathbf{M}\Bigl(T_{\mathrm{S},\mathbf{M}}q^{2\kappa S+2\al S(0)}\Bigr)}
\,.
\label{Zkappa}
\end{align}
Here $q=e^{\pi i\nu}$ is related to the coupling parameter
(see \eqref{q-nu} below), 
and $T_{S,\mathbf{M}}$ stands for
the monodromy matrix on the two 
tensor products of evaluation representations 
of $U_q(\slth)$: 
one for the horizontal (or `space') direction $\mathrm{S}$, 
and another for the vertical (or Matsubara) 
direction $\mathbf{M}$. 
For more details see section \ref{XXZ} below,  
in particular {\it fig. 1}.  
It was important in \cite{HGSIII}  
to incorporate inhomogeneities in the Matsubara chain. 
This allows, for example, to consider the 
temperature expectation values 
in the spirit of \cite{Klumper,Suzuki}, 
by adjusting inhomogeneities and  
taking the limit to the infinite chain 
in the Matsubara  direction. 

The clue to our calculation was the introduction 
of operators $\tb ^*(\z)$, $\bb ^*(\z)$, $\cb ^*(\z)$ 
\cite{HGSII} which, 
by acting on the primary field $q^{2\al S(0)}$,  
create the space of quasi-local operators on the 
horizontal chain.  
More precisely, quasi-local operators are created 
by Taylor coefficients of  $\tb ^*(\z)$, $\bb ^*(\z)$, 
$\cb ^*(\z)$ at the point $\z^2=1$.
In this paper we change the definition of 
$\bb ^*(\z)$, $\cb ^*(\z)$ 
from those of \cite{HGSII,HGSIII} by applying certain Bogolubov
transformation.
Compared with the original ones, they have better asymptotic properties. 
We shall explain this in section \ref{XXZ}.

There is an obvious similarity with conformal 
field theory (CFT), where 
the descendants are created from the primary
field by the action of the Virasoro algebra. 
Our aim in this paper is to 
examine the scaling limit of our construction in the critical regime,   
and to establish its precise relation with CFT. 
We shall consider the case of a homogeneous Matsubara chain.

The functional \eqref{Zkappa}
is non-trivial only for 
operators $\mathcal{O}$ of spin zero. 
As it turns out, for the study of the scaling limit,  
it is quite useful to relax this restriction. 
We shall first introduce the following generalisation of \eqref{Zkappa}
which is of interest on its own right. For $s>0$ we define
$$
Z^{\kappa, s}
\Bigl\{q^{2\al S(0)}\mathcal{O}
\Bigr\}=
\frac{\Tr _{\mathrm{S}}\Tr _{\mathbf{M}}\Bigl(Y_\mathbf{M}^{(-s)}T_{\mathrm{S},\mathbf{M}}\ q^{2\kappa S}
\ \mathbf{b}^*_{\infty,s-1}\cdots \mathbf{b}^*_{\infty, 0}
\bigl(
q^{2 \al S(0)}\mathcal{O}\bigr)\Bigr)}
{\Tr _{\mathrm{S}}\Tr _{\mathbf{M}}\Bigl(Y_\mathbf{M}^{(-s)}T_{\mathrm{S},\mathbf{M}}\ q^{2\kappa S}
\ \mathbf{b}^*_{\infty,s-1}\cdots \mathbf{b}^*_{\infty, 0}
\bigl(
q^{2 \al S(0)}\bigr)\Bigr)}
\,,
$$
where the $\mathbf{b}^*_{\infty,j}$'s denote 
the coefficients of the singular part of $\bb^*(\z)$ at $\z^2=0$. 
When $s<0$, a similar definition is in force using 
the expansion coefficients of $\cb^*(\z)$. 
As long as $Y_\mathbf{M}^{(-s)}$ 
is taken generically, this definition is independent of its choice 
(see section \ref{XXZ} for more details). 
In general the dependence on $Y_\mathbf{M}^{(-s)}$ 
enters, but only in a ``topological" way. 

Needless to say, what we 
we are dealing with is a lattice analogue of 
the screening operators {\` a} la Feigin-Fuchs-Dotsenko-Fateev \cite{FF,DF}. 
It is interesting to see that 
quasi-local operators and screening operators  both arise
from the same operators $\bb^*(\z)$, $\cb^*(\z)$, 
as expansions either around $\z ^2=1$ or $\z ^2=0$.

After these modifications the main formula of \cite{HGSIII} 
remains valid.
It reads  
\begin{align}
&Z^{\kappa,s}\bigl\{\tb^*(\z^0_1)\cdots \tb^*(\z^0_p)
\bb^*(\z^+_1)\cdots \bb^*(\z^+_r)
\cb^*(\z^-_r)\cdots \cb^*(\z^-_1)\bigl(q^{2\al  S(0)}\bigr)\bigr\}
\label{mainZ}
\\
&=\prod\limits _{i=1}^p 2\rho (\z _i^{0}|\kappa,\kappa+\al,s)\times
\det \left(\omega(\z^+_i,\z ^-_j|\kappa,\al,s)
\right)_{i,j=1,\cdots, r}\,,\nn
\end{align}
where the functions $\rho (\z _i^{0}|\kappa,\kappa+\al,s)$ and 
$\omega(\z^+_i,\z ^-_j|\kappa,\al,s)$ are defined by the data of the Matsubara  direction.
We refer to this formula as  the determinant formula. 
Due to the Bogolubov transformation of
$\bb^*$ and 
$\cb^*$ the function $\omega(\z,\xi|\kappa,\al,0)$ is slightly
different from $\omega(\z,\xi |\kappa,\al)$ used in \cite{HGSIII,BG}.

Now let us turn to the scaling limit. 
We have two twisted transfer matrices 
$T_\mathbf{M}(\z,\kappa +\al)$ and $T_\mathbf{M}(\z,\kappa)$
in the Matsubara direction. 
As the the number of sites $\mathbf{n}$ becomes large, 
their Bethe roots tend to distribute densely on $\R_+$. 
Fixing $R>0$ and introducing the step of the lattice $a$,  
we consider the limit  
\begin{align}
\mathbf{n}\to\infty,\quad a\to 0,\quad \mathbf{n}a=2\pi R \ \ \mathrm{fixed}\,.
\label{scaling1}
\end{align}
At the same time we rescale the spectral parameter as
\begin{align}
\z =(C a)^\nu \la, \quad \la \ \ \mathrm{fixed}\,,
\label{scaling2}
\end{align}
so that 
the Bethe roots close to $0$ stay finite
in terms of the variable $\lambda$.
Here $C$ is a constant 
chosen for fine tuning 
(see section \ref{contmatsubara}, \eqref{C}). 
In this limit the twisted transfer matrices turn into
the transfer matrices of chiral CFT on the cylinder 
$Cyl=\mathbb{C}/2\pi i R\,\mathbb{Z}$,  
introduced and studied 
by Bazhanov, Lukyanov and Zamolodchikov \cite{BLZI,BLZII}.
We wish to mention here that the present work owes a great deal 
to these remarkable papers without 
which it would have been impossible. 
Details about the scaling limit can be found 
in section \ref{contmatsubara} below.  
The relevant CFT has central charge 
$$
c=1-\frac {6\nu^2}{1-\nu}\,.
$$
We shall parametrise the conformal dimension as 
\begin{align*}
\Delta_{\alpha}=\frac{\nu^2}{4(1-\nu)}
\bigl((\alpha-1)^2-1\bigr),
\end{align*}
and 
write the action of the Virasoro algebra
on a local field $\phi(y)$ 
as $\mathbf{l}_{n}(\phi)(y)$. 

In the following we set $\bar{a}=C a$, and let 
$\lim_\mathrm{scaling}$ indicate the scaling limit  
\eqref{scaling1}, \eqref{scaling2}.  
The functions entering the determinant formula 
\eqref{mainZ} also have finite limits, 
\begin{align}
&\rhob(\la|\kappa,\kappa')
=\lim_\mathrm{scaling}\ 
\rho(\la \bar{a}^{\nu}|\kappa,\al,s)\,,
\nn\\
&4\ \omegab(\la,\mu|\kappa,\kappa ',\al)
= \lim_\mathrm{scaling}\ 
\omega(\la \bar{a}^\nu,\mu 
\bar{a}^\nu|\kappa,\al,s)\,, 
\nn
\end{align}
where 
\begin{align}
\kappa'=\kappa +\al+2\textstyle {\frac {1-\nu}\nu}s\,. 
\label{kappa-prime}
\end{align}
So all these partition functions \eqref{mainZ} 
have finite limits. 
They should have some  
definite meaning in the context of CFT. 
We contend that they are the three point functions 
of the descendants of the chiral primary field 
$\phi_{\al}(0)$, computed in the presence of 
two other primary fields (or their descendants)  
inserted at the two ends of the cylinder. 

More specifically, we conjecture that the following 
picture holds true. 
First, the creation operators tend to a limit, 
\begin{align}
2\taub^*(\la)=\lim_\mathrm{scaling}\tb^*(\la \bar{a}^{\nu}),\ \ 
2\betab^*(\la)=\lim_\mathrm{scaling}\bb^*(\la \bar{a}^{\nu}),\ \ 
2\gammab^*(\la)=\lim_\mathrm{scaling}
\cb^*(\la \bar{a}^{\nu})\nn\,.
\end{align}
As $\lambda\to\infty$, 
these operators have asymptotic expansions of the form
\begin{align}
&\log\(\taub^*(\la)\)\simeq
\sum\limits _{j=1}^{\infty}
\taub^*_{2j-1}\la^{-\frac{2j-1}\nu}
\,,
\label{scale3}\\
&\frac 1{\sqrt{\taub^*(\la)}}\betab^*(\la)
\simeq\sum\limits _{j=1}^{\infty}
\betab^*_{2j-1}
\la^{-\frac{2j-1}\nu}
\,,
\quad\frac 1{\sqrt{\taub^*(\la)}}\gammab^*(\la)
\simeq
\sum\limits _{j=1}^{\infty}\gammab^*_{2j-1}\la^{-\frac{2j-1}\nu}
\,.\nn
\end{align}
In the limit, the quasi-local operator $q^{2\al S(0)}$ 
becomes the product of two chiral primary fields 
$\phi_{\al}(0)\otimes \bar{\phi}_{-\al} (0)$.  
The operators $\taub^*_{2j-1}$ and the quadratic 
combinations $\betab^*_{2j-1}\gammab^*_{2k-1}$ 
act only on the left component $\phi_{\al}(0)$, 
and create the entire Verma module 
spanned by the Virasoro descendants
\begin{align}
\mathbf{l}_{-m_1}\cdots \mathbf{l}_{-m_s}(\phi _{\al})(0)
\,.
\nn
\end{align}
Furthermore, if $Y^{(-s)}_\mathbf{M}$ is chosen to be generic 
then
\begin{align}
&\lim_\mathrm{scaling}
Z^{\kappa,s}
\bigl\{\tb^*(\z^0_1)\cdots \tb^*(\z^0_p)
\bb^*(\z^+_1)\cdots \bb^*(\z^+_r)
\cb^*(\z^-_r)\cdots \cb^*(\z^-_1)
\bigl(q^{2\al S(0)}\bigr)\bigr\}
\label{scaling4}\\
&=
2^{p+2r}Z_{R}^{\kappa,\kappa '}
\bigl\{\taub^*(\la^0_1)\cdots \taub^*(\la^0_p)
\betab^*(\la^+_1)\cdots \betab^*(\la^+_r)
\gammab^*(\la^-_r)\cdots \gammab^*(\la^-_1)
\bigl(\phi_{\al}(0)\bigr)\bigr\}\,.
\nn
\end{align}
In the right hand side, the symbol 
$Z_{R}^{\kappa,\kappa '}\{X(0)\}$ 
stands for the three point function 
normalised as $Z_{R}^{\kappa,\kappa '}\{\phi_\al(0)\}=1$, 
with $X(0)$ inserted at $x=0$ and 
the primary fields $\phi_{\kappa +1}$, 
$\phi_{-\kappa' +1}$ being 
inserted at $x=\infty$ and $x=-\infty$, respectively  
(see \eqref{ZR} below). 
Non-generic choice of $Y^{(-s)}_\mathbf{M}$ 
corresponds to replacing the primary fields 
at $x=\pm\infty$ by their descendants.  
In this paper we discuss only the case of 
generic $Y^{(-s)}_\mathbf{M}$. 

The coefficients in \eqref{scale3} are homogeneous operators
in the sense that 
$$
\left[\ \mathbf{l}_0\,, \taub^*_{2j-1}\right]
=(2j-1)\taub^*_{2j-1},
\quad 
\left[\ \mathbf{l}_0\,, \betab^*_{2i-1}\gammab^*_{2j-1}\right]
=(2i+2j-2)\betab^*_{2i-1}  \gammab^*_{2j-1} \,. 
$$
Hence, for each degree, the descendants created by 
$\mathbf{l}_{-k}$'s, 
and those created by $\taub^*_{2j-1}$'s,
$\betab^*_{2j-1}$'s and  $\gammab^*_{2j-1}$'s, 
must be finite linear combinations of each other. 
The main goal of this paper is to 
show, for low degrees, 
that this is indeed the case, and that the coefficients 
can be found explicitly.   

To determine the coefficients of the linear combination, 
we compare the values of $Z_{R}^{\kappa,\kappa'}$. 
For the Virasoro descendants, they can be easily 
computed by the conformal Ward-Takahashi identities. 
For the descendants by $\taub^*_{2j-1}$ and others, 
we need the coefficients of the 
asymptotic expansion of the functions 
$\rho _R(\la|\kappa,\kappa ')$ and 
$\omega_R(\la,\mu|\kappa,\kappa ',\al)$ .
In section \ref{asymp-loga} 
we develop a systematic method for computing them.  
We note that in both cases the results 
are polynomials in the conformal dimensions 
$\Delta_{\kappa+1}$, $\Delta_{\kappa'+1}$.   
We may regard them as independent variables and 
compare the coefficients,  
since $s$ in \eqref{kappa-prime} can take any integer values.  
This was one of reasons for us to introduce 
the screening operators. 

We consider first $\taub^*_{2j-1}$. 
CFT allows an integrable structure  
based on Zamolodchikov's integrals of motion 
$\mathbf{i}_{2m-1}$ \cite{Zam}.
With the above procedure we are led 
to a result which should not be surprising, 
\begin{align*}
\taub^*_{2m-1}=C_m\cdot \mathbf{i}_{2m-1}\,,
\end{align*}
where $C_m$ are some 
$\nu$-dependent constants which can be found in \cite{BLZII}.

We then consider the action of 
$\betab^*_{2j-1}$'s and  $\gammab^*_{2j-1}$'s. 
Because of a technical difficulty we have not been able to 
compute the asymptotics of 
$\omega_R(\la,\mu|\kappa,\kappa ',\al)$
for $\kappa\neq\kappa'$. 
Here we restrict to the case $\kappa=\kappa'$. 
Since $Z^{\kappa,\kappa}_R(\mathbf{i}_{2n-1}(X))=0$ for any 
$X$, restricting to $\kappa=\kappa'$ means that we consider 
the quotient space of the Verma module modulo the action of 
the integrals of motion. 
We assume that the vectors
$$
\mathbf{i}_{2k_1-1}\cdots \mathbf{i}_{2k_r-1}
\mathbf{l}_{-2m_1}\cdots \mathbf{l}_{-2m_s}(\phi _{\al}(0))
$$
span the Verma module, so 
the quotient space is created by the $\mathbf{l}_{-2m}$'s. 
With primary fields as asymptotical states, we 
can compare up to the level 6. 
It should be added, however, that up to this level 
the system of equations is overdetermined. 
So the very possibility of finding a solution is the 
strongest support of our fermionic picture. 

We give one example on the level 4:
$$
\betab ^*_1\gammab^*_3(\phi_\al(0))
=\textstyle{\frac 1 2}D _1(\al)D _3(2-\al)
\Bigl(\mathbf{l}_{-2}^2 
+\frac {2c-32 -6d_{\al}} 9 \mathbf{l}_{-4}\Bigr)(\phi_\al(0))\,,
$$
where
\begin{align}
&d_\alpha=
\textstyle{\frac 1 6}
\sqrt{(25-c)(24\Delta _{\al}+1-c)}\,,
\nn\\     
&D_{2n-1}(\alpha)=
\frac 1 {\sqrt{i\nu}}
\ \Gamma (\nu)^{-\frac{2n-1}\nu}(1-\nu)^{\frac{2n-1}{2}}
\cdot \frac{1}{(n-1)!}
\frac{\Gamma\left(\frac{\alpha}{2}+\frac 1{2\nu}(2n-1)\right)}
{\Gamma\left(\frac{\alpha}{2}+\frac{(1-\nu)
}{2\nu}(2n-1)
\right)}
\,.\nn
\end{align}
In the right hand side, we have 
a particular combination of the Virasoro descendants. 
This equation says that 
its three-point function remains of the same 
determinant form before and after integrable perturbation.

The text is organised as follows. 
In section \ref{XXZ} we review the results of 
\cite{HGSIII} and describe the Bogolubov transformation
mentioned above.
In section \ref{rhomega} we define the functions $\rho$ and $\omega$.
In section \ref{screenings} we define screening operators
on the lattice, and describe a generalisation of the previous results.
In section \ref{scaling} we start discussing the scaling 
limit of the XXZ chain, examining the behaviour of 
the Bethe roots in the Matsubara direction
as the length of the chain becomes infinite. 
Sections \ref{CFT} and \ref{reviewBLZ}
are a review of the CFT 
integrals of motion on the cylinder, and 
the series of works of 
Bazhanov, Lukyanov and Zamolodchikov (BLZ).  
We explain in section \ref{contmatsubara}
how the Matsubara transfer matrix turns into that of BLZ
in the continuous limit.  In section \ref{cftlimitspace}
we discuss the CFT
interpretation of the scaling limit in the space direction.
In section  \ref{asymp-loga} we study the asymptotics of 
Thermodynamic Bethe Ansatz (TBA) function $\frak{a}$ for CFT.
In section \ref{asymp-omega} 
we find the asymptotical expansion of $\omega$
for $\kappa=\kappa '$. In section \ref{final} we compare descendants created by $\mathbf{l}_{-2m}$ with
those created by $\betab^*_{2j-1}$'s and  $\gammab^*_{2j-1}$ and give some concluding remarks.
In appendix we present general properties of asymptotics of $\omega$ which apply to the
case $\kappa \ne\kappa '$.

\section{Review of previous results}\label{XXZ}

Let us start with a brief review of the papers 
\cite{HGSI,HGSII,HGSIII}.
Consider the XXZ spin chain in the infinite volume. 
The space of states of the model is 
$$
\frak{H}_\mathrm{S} 
=\bigotimes\limits_{j=-\infty}^{\infty}\mathbb{C}^2\,,
$$ 
and the Hamiltonian is given by
\begin{eqnarray}
H=\textstyle{\frac{1}{2}}\sum\limits_{k=-\infty}^{\infty}
\left( 
\sigma_{k}^1\sigma_{k+1}^1+
\sigma_{k}^2\sigma_{k+1}^2+
\Delta\sigma_{k}^3\sigma_{k+1}^3
\right), \quad \Delta =\half(q+q^{-1})\,.
\label{Ham}
\end{eqnarray}
We consider the critical XXZ model in the following range of the coupling, 
\begin{equation}
q=e^{\pi i\nu},\quad 1/2< \nu <1 \,.
\label{q-nu}
\end{equation}

Together with $\frak{H}_\mathrm{S} $ we consider 
the Matsubara space 
$\frak{H}_{\mathbf{M}}$. In \cite{HGSIII} 
the most general case was treated:  
namely, $\frak{H}_{\mathbf{M}}$ was the tensor product 
of spaces of different dimensions, and to every site $\mathbf{m}$ 
an independent inhomogeneity parameter 
$\tau _\mathbf{m}$ was attached.  
In the present paper
we shall restrict ourselves to the case
$$
\frak{H}_{\mathbf{M}}=
\bigotimes\limits_{\mathbf{j=1}}^{\mathbf{n}}\mathbb{C}^2\,.
$$ 

We consider the monodromy matrix $T_{\mathrm{S},\mathbf{M}}$. 
Mathematically this is nothing but the image of 
the universal $R$ matrix of $U_q(\slth)$ 
on the tensor product of evaluation representations 
corresponding to $\frak{H}_{\mathrm{S}}$ 
and $\frak{H}_{\mathbf{M}}$. It has been said that we shall 
consider a homogeneous Matsubara chain only. In the notation
of \cite{HGSIII},  
this correspond to setting $\tau _\mathbf{m}=q^{\frac 1 2}$ 
for all $\mathbf{m}$. We shall 
absorb this into redefinition
of the $L$-operator comparing to \cite{HGSIII}. Let us write the definition explicitly:
$$
T_{\mathrm{S},\mathbf{M}}=\raisebox{.7cm}{$\curvearrowright $} 
\hskip -.75cm\prod\limits_{j=-\infty}^{\infty}
T_{j,\mathbf{M}} \,,
$$
where
$$
T_{j,\mathbf{M}}\equiv T_{j,\mathbf{M}}(1),
\quad T_{j,\mathbf{M}}(\z)=\raisebox{.7cm}{$\curvearrowleft $} 
\hskip -.6cm\prod\limits_{\mathbf{m=1}}^{\mathbf{n}}
L_{j,\mathbf{m}}(\z)
\,,
$$
with
$$
L_{j,\mathbf{m}}(\zeta)
=q^{-\frac 1 2\sigma ^3_j\sigma ^3_\mathbf{m}}
-\z ^2 q^{\frac 1 2\sigma ^3_j\sigma ^3_\mathbf{m}}
-\z (q-q^{-1}) (\sigma ^+_j\sigma ^-_\mathbf{m}+\sigma ^-_j\sigma ^+_\mathbf{m})\,.
$$ 

A local operator $\mathcal{O}$ on $\frak{H}_\mathrm{S}$ 
is by definition
an operator which acts nontrivially only on a finite number of 
the tensor components $\mathbb{C}^2$ of $\frak{H}_\mathrm{S}$. 
More generally we consider quasi-local operators 
with tail $\al$, which are operators of the form   
$$
q^{2\al S(0)}\mathcal{O},\quad 
S(k)=\half\sum\limits _{j=-\infty}^k\sigma ^3_j\,,
$$
with $\mathcal{O}$ being local. 
In this notation  $S=S(\infty)$ is the total spin. 
In \cite{HGSIII} we computed the expectation values defined by 
\begin{align}
Z^{\kappa}\Bigl\{q^{2\al S(0)}\mathcal{O}
\Bigr\}=
\frac{\Tr _{\mathrm{S}}\Tr _{\mathbf{M}}\Bigl(T_{\mathrm{S},\mathbf{M}}q^{2\kappa S+2\al S(0)}\mathcal{O}\Bigr)}
{\Tr _{\mathrm{S}}\Tr _\mathbf{M}\Bigl(T_{\mathrm{S},\mathbf{M}}q^{2\kappa S+2\al S(0)}\Bigr)}
\,.
\label{Z}
\end{align}
This is a linear functional on the space $\mathcal{W}_{\al,0}$ of spinless 
quasi-local operators with tail $\al$.

It is helpful to think of the functional $Z^{\kappa}$ 
as a ratio of partition functions of the six vertex model
on the cylinder. 
For example, the numerator of \eqref{Z} can be presented graphically as follows.

\vskip .5cm
\hskip .5cm\includegraphics[height=7cm]{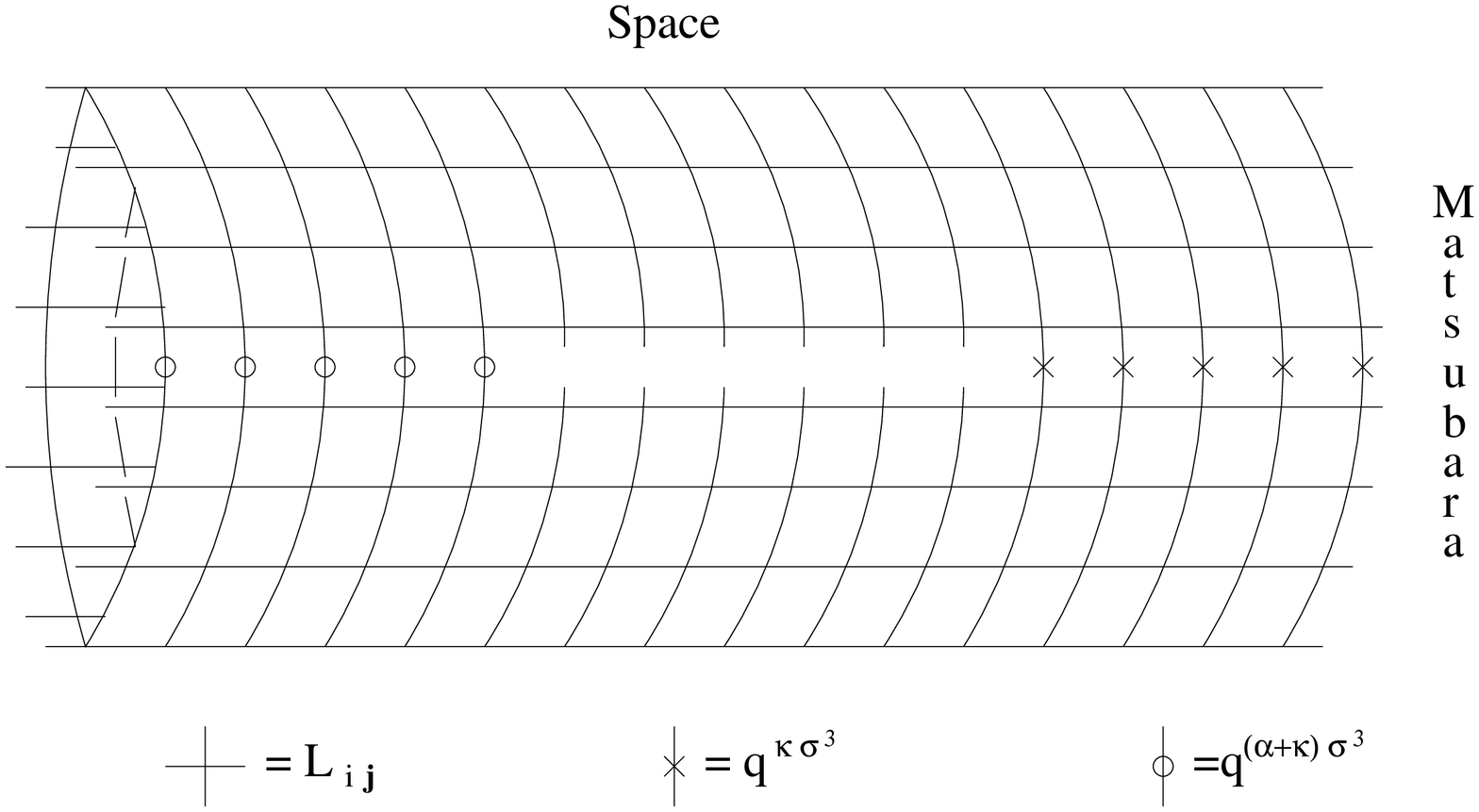}
\vskip .5cm
{\it Fig.1: Partition function on the cylinder. \\

The functional $Z^\kappa$ is a ratio of two partition functions on the cylinder. 
On each crossing of a row and a column, one associates the Boltzmann weights of the 
six vertex model. On a particular row there are also 
twist fields $q^{\kappa \sigma^3_j}$ (marked by crosses)
and  $q^{(\alpha+\kappa) \sigma^3_j}$ (marked by circles).
The numerator of $Z^\kappa$ corresponds to a lattice with 
defects representing an insertion of a local operator.
}

\vskip .5cm

\vskip 2cm
On this picture the summation is performed over all edges except the broken ones in the middle.
The arrows on the broken edges are fixed, 
representing the local operator $\mathcal{O}$. 
The one dimensional sublattice 
going in the infinite space direction will be called 
the space chain,
while the compact one dimensional sublattice in the 
Matsubara direction will be referred to as the Matsubara chain.

We computed the expectation values \eqref{Z} using the fermionic 
description of $\mathcal{W}_{\al,0}$ found in \cite{HGSII}. 
Let us briefly recall it. 
Consider the space
$$
\mathcal{W}^{(\al)}
=\bigoplus\limits _{s=-\infty}^{\infty}\mathcal{W}_{\al-s,s}
\,,
$$
where $\mathcal{W}_{\al-s,s}$ denotes
 the space of quasi-local operators of
spin $s$ with tail $\al -s$.

We have the creation operators $\tb ^*(\z)$, $\bb^*(\z)$, $\cb^*(\z)$ 
and the annihilation operators $\bb(\z)$, $\cb(\z)$
which act on $\mathcal{W}^{(\al)}$. To be precise the operators 
$\tb ^*(\z)$, $\bb^*(\z)$, $\cb^*(\z)$ were defined in \cite{HGSII}
as formal power series in $\z^2-1$, the quasi-local operators
in $\mathcal{W}^{(\al)}$ are created by coefficients of these series.
However, when the series are substituted into $Z^{\kappa}$ the result
allows analytical continuation. So, in the present paper we shall adopt another point of view
which is similar to that of CFT.
Namely, we shall consider the operators $\tb ^*(\z)$, $\bb^*(\z)$, $\cb^*(\z)$ as
analytical functions.
Then the relation to real quasi-local operators is achieved by considering 
$\tb ^*(\z)$, $\bb^*(\z)$, $\cb^*(\z)$ around the point $\z^2=1$.

The creation-annihilation operators have  the block structures
\begin{align}
&\tb ^*(\z) \ \ \ \ 
\ \ \ \ :\ \mathcal{W}_{\al
-s,s}\ \to\ \mathcal{W}_{\al-s,s}\label{blocks}\\
&\bb ^*(\z) ,\cb (\z)
\ :\ \mathcal{W}_{\al-s+1,s-1}\ \to\ \mathcal{W}_{\al-s,s}\,,\nn\\
& \cb ^*(\z),
\bb(\z)
\ :\ \mathcal{W}_{\al-s-1,s+1}\ \to\ \mathcal{W}_{\al-s,s}\,.\nn
\end{align}
The operator $\taub=\tb^*_1/2$ plays a special role:
it is the right shift by one site along the space chain. 

We have
\begin{align}
&\bb(\z)(q^{2\al S(0)})=0,\qquad\quad\quad\ \ \  \cb(\z)(q^{2\al S(0)})=0\,,\nn\\
&\bigl[  \cb(\xi),\cb^*(\z)  \bigr]_+=\psi (\xi/\z,\al),\quad 
\bigl[  \bb(\xi),\bb^*(\z)  \bigr]_+=-\psi (\z/\xi,\al)\,,\nn
\end{align}
where 
\begin{align}
\psi (\z,\al)=\z^{\al}\frac{\z^2 +1}{2(\z^2-1)}\,.\label{defpsi}
\end{align}
 The operators in the space $\mathcal{W}_{\al,0}$
are created from the primary field $q^{2\al S(0)}$ by
action of $\tb ^*$'s and of equal number of $\bb ^*$'s and
$\cb ^*$'s. 
The completeness \cite{complet} says that the entire space
${\mathcal{W}}_{\al,0}$ is generated  by coefficients of 
the creation operators considered as series in $\z ^2-1$.
Certainly, this description is reminiscent of CFT.

The main result of \cite{HGSIII} is the following relations 
which allow for recursive computations of the expectation values: 
\begin{align}
&Z^{\kappa}\bigl\{\tb^*(\z)(X)\bigr\}
=2\rho(\z|\kappa,\kappa+\al)Z^{\kappa}\{X\}\,,\label{maint}\\
&Z^{\kappa}\bigl\{\bb^*(\z)(X)\bigr\}
=\frac 1{2\pi i}\oint\limits _{\Gamma}
\omega (\z,\xi|\kappa,\al)
Z^{\kappa}\bigl\{\cb(\xi)(X)\bigr\}
\frac{d\xi^2}{\xi^2}\,,\label{mainb}\\
&Z^{\kappa}\bigl\{\cb^*(\z)(X)\bigr\}
=-\frac 1 {2\pi i}\oint\limits_{\Gamma}
\omega (\xi,\z|\kappa,\al)
Z^{\kappa}\bigl\{\bb(\xi)(X)\bigr\}
\frac{d\xi^2}{\xi^2}\,,
\label{mainc}
\end{align}
where $\Gamma$  goes around all the singularities of
the integrand except $\xi^2=\z^2$. We think no further
explanation is needed  nowadays when the method
of CFT is a part of common knowledge.

The functions $\rho (\z|\kappa,\kappa +\al) $ 
and $\omega(\zeta,\xi|\kappa,\alpha)$ will be defined in 
section \ref{rhomega}.  
We changed the notation for the former from to \cite{HGSIII}.
The present notation
agrees better with the explicit formula given below.
The set of equations \eqref{maint},  \eqref{mainb}, \eqref{mainc} implies
a determinant representation for the expectation values \cite{HGSIII}, we shall discuss this later. 

Let us describe the modification of 
$\bb ^*(\z)$, $\cb ^*(\z)$ by a Bogolubov transformation 
which was mentioned in the introduction.   
Denoting the operators used in \cite{HGSII,HGSIII} 
by $\bb _\mathrm{rat}^*(\z)$, 
$\cb_\mathrm{rat} ^*(\z)$, 
define
\begin{align}
&\bb ^*(\z)=\bb _\mathrm{rat}^*(\z)+\frac 1{2\pi i}\int\limits _{\Gamma}
D_\z D_\xi\Delta ^{-1}_\z\psi (\z/\xi,\al)\cdot 
\cb (\xi)\frac {d\xi ^2}{\xi ^2}\,,\label{modtbc}\\
&\cb ^*(\z)=\cb_\mathrm{rat} ^*(\z)-\frac 1{2\pi i}\int\limits _{\Gamma}
D_\z D_\xi\Delta ^{-1}_\z\psi (\xi/\z,\al)\cdot 
\bb (\xi)\frac {d\xi ^2}{\xi ^2}\,.\nn
\end{align}
where
$D_\z$ is the following finite difference operator of the 
second order, 
$$
D_\z f(\z)=f(\z q)+f(\z q^{-1})-\tb^*(\z)f(\z)\,.
$$
The function $\Delta ^{-1}_\z\psi (\z,\al)$ is transcendental. 
For 
\begin{align*}
\z^2>0, \quad 
-\frac 1 \nu<\mathop{{\rm Re}}\al<0
\end{align*}
we define it as 
\begin{align}
\Delta ^{-1}_\z\psi (\z,\al)=-VP\int\limits _0^{\infty}\frac 1{2\nu\bigl(1+\(\z/\eta\)^{\frac 1 \nu}\bigr)}
\psi (\eta,\al)\frac{d\eta ^2}{2\pi i\eta ^2}\,,
\label{Dinv} 
\end{align}
where the principal value is taken 
with regards to the pole at $\eta^2=1$. 
In general we define it by analytic
continuation with respect to both $\al$ and $\z ^2$, 
obtaining a meromorphic function of $\log\z$.   
It is bounded at $\log \z \to\pm\infty$, and  
its singularities closest to the real axis are 
the simple poles at $\log \z =\pm \pi i \nu$ 
with residues of opposite signs. 

The function 
$D_\z D_\xi\Delta ^{-1}_\z\psi (\z/\xi,\al)$ 
is regular at $\z=\xi$, so the Taylor series for $\bb ^*(\z)$, $\cb ^*(\z)$
at $\z^2=1$ are well-defined. The function $\omega$ changes
following the change of $\bb ^*$, $\cb^*$:
\begin{align}
\omega (\z,\xi|\kappa,\al)=
\omega _\mathrm{rat}(\z,\xi|\kappa,\al)+\overline{D}_\z
\overline{D}_\xi\Delta ^{-1}_\z\psi (\z/\xi,\al)\,,
\label{omega-tilde}
\end{align}
where 
\begin{align}
\overline{D}_\z f(\z)=f(\z q)+f(\z q^{-1})-2\rho (\z |\kappa,\kappa+\al)f(\z)\,.\label{Dbar}
\end{align}

Let us mention one marvellous property of our modified
operators $\bb^*$, $\cb^*$.

The main subject of our original study
\cite{HGSI,HGSII} was the following normalised
matrix element 
\begin{align}
Z_{\infty}\{q^{2\al S(0)}\mathcal{O}\}
=\frac{\langle \mathrm{vac}|q^{2\al S(0)}
\mathcal{O}|\mathrm{vac}\rangle}
{\langle \mathrm{vac}|q^{2\al S(0)}|\mathrm{vac}\rangle}\,,
\label{Zinf}
\end{align}
where $|\mathrm{vac}\rangle$ denotes the 
ground state of the XXZ Hamiltonian in the infinite volume. 
Equivalently, the numerator of
\eqref{Zinf} is the partition function 
of the six vertex model on the
plane with a defect localised between two horizontal lines.  

It is easy to see using the formulae from 
\cite{HGSI,HGSII} that,  
if we create the quasi-local fields by 
the operators $\tb ^*(\z)$, 
$\bb ^*(\z)$, $\cb ^*(\z)$,  
then $Z_{\infty}$ vanishes on all of them
with the sole exception of the descendants 
created by $\taub=\tb^*_1/2$. 
For the latter we have
$$
Z_{\infty}\{\taub^m(q^{2\al S(0)}\mathcal{O}) \}=1,
\quad m\in \mathbb{Z}\,.
$$

We know from the algebraic construction \cite{HGSII}
that the Taylor coefficients of the part
 $\bb ^*_\mathrm{rat}(\z)$, $\cb ^*_\mathrm{rat}(\z)$ 
in \eqref{modtbc}
produce only rational functions of $q$ and $q^\alpha$. 
Hence all transcendental pieces of the 
expectation values \eqref{modtbc} 
come from rewriting a given quasi-local operator in 
the fermionic basis and picking Taylor coefficients of 
the function $\overline{D}_\z\overline{D}_\xi 
\Delta ^{-1}_\z \psi (\z/\xi,\al)$.

The property of 
$\tb ^*(\z)$ (except $\tb^*_1$) and 
$\bb ^*(\z)$, $\cb ^*(\z)$ mentioned above is analogous to 
that of the Virasoro generators in CFT: 
on the Riemann sphere,  
all normalised one point functions of the descendants 
vanish due to the conformal invariance.  
Strictly speaking, the one point function 
of the primary field also vanishes, so a word of 
clarification is necessary.  
Take a massive model with a mass scale $m$  
and consider the conformal limit $m\to 0$. 
While the normalised one point
function of the primary field stays equal to $1$,   
those of the descendants vanish in the limit,  
because for dimensional reasons 
they contain additional powers of $m$.

At this point one may wonder why we 
did not do the Bogolubov
transformation killing completely the 
function $\omega$ in \eqref{mainb}, \eqref{mainc}.  
The answer is that it is impossible 
to rewrite the left hand sides of \eqref{mainb}, \eqref{mainc} 
because the function $\omega$ 
cannot be written as a function of $\rho$.

\section{ Functions $\rho$ and $\omega$}\label{rhomega}

Introduce the twisted Matsubara transfer matrix:
\begin{align}
&T_{\mathbf{M}}(\z ,\kappa)
=\Tr _j\bigl(T_{j,\mathbf{M}}(\z)q^{\kappa\sigma ^3_j}\bigr)
\,. \label{T0}
\end{align}
Let  $|\kappa\rangle$ be the eigenvector of $T(1,\kappa)$ 
whose eigenvalue is maximal in the absolute value. Similarly let 
$\langle \kappa +\al|$ be the eigencovector of $T(1,\kappa+\al)$ 
whose eigenvalue is maximal in the absolute value. 
It is well-known that these eigenvectors have spin $0$. 
We call them the maximal eigenvectors,   
and assume that they are not orthogonal.
We denote the eigenvalues of 
$T_{\bf M}(\zeta,\kappa)$ (resp. $T_{\bf M}(\zeta,\kappa+\alpha)$)
on $|\kappa\rangle$ (resp. $\langle \kappa +\al|$) 
by $T(\zeta,\kappa)$ (resp. $T(\zeta,\kappa+\alpha)$).

Then $\rho (\z|\kappa,\kappa +\al)$ is defined by 
$T(\z,\kappa)$ and 
$T(\z,\kappa +\al)$. We have
\begin{align}
\rho(\z|\kappa, \kappa +\al)=\frac{T(\z,\kappa +\al)}{T(\z,\kappa)}\,.
\label{badrho}
\end{align}
The function $\omega(\z,\xi|\kappa,\al)$ is more complicated.
In \cite{HGSIII} it was shown that it is completely determined
by two requirement: the singular part  and the normalisation condition.
Then they were explicitly solved in terms $q$-deformed Abelian integrals.
In the present paper it is convenient to use an alternative,
TBA-like,  description
used in \cite{BG}. Let us explain this simplifying a little the
notation of \cite{BG}.

Together with the transfer matrix $T_\mathbf{M}(\z,\kappa)$
we consider in \cite{HGSIII} Baxter's $Q$-operators  $Q^{\pm}_\mathbf{M}(\z,\kappa)$. In this paper we  use only
one of them: $Q^{-}_\mathbf{M}(\z,\kappa)$ denoting it just as
$Q_\mathbf{M}(\z,\kappa)$. 
It is defined by 
the trace over the highest weight representation of
the $q$-oscillator algebra with generators $\mathbf{a}$,
$\mathbf{a}^*$, $D$ (see \cite{HGSII} for notation), 
\begin{align}
&Q_{\mathbf{M}}(\z ,\kappa)
=\z ^{-\kappa+S_\mathbf{M}}
(1-q^{-2(\kappa-S_\mathbf{M})})
\Tr^- \bigl(T^-_{Osc,\mathbf{M}}(\z)q^{-2\kappa D_A}\bigr)
\,, \label{Q0}
\end{align}
where 
\begin{align}
&  T^-_{Osc,\mathbf{M}}(\z)=  \raisebox{.7cm}{$\curvearrowleft $} 
\hskip -.6cm\prod\limits_{\mathbf{m=1}}^{\mathbf{n}}
L^-_{Osc,\mathbf{m}}(\z)\,,\nn\\
&L^-_{Osc,\mathbf{m}}(\z)=\bigl(I-\z ^2 q^{2D+1}
\sigma ^-_\mathbf{m}\sigma^+ _\mathbf{m}-\z q^{-\frac 1 2}
(\mathbf{a}\ \sigma _\mathbf{m}^-+\mathbf{a} ^*\sigma _\mathbf{m}^+)\bigr)
  q^{D\sigma ^3_\mathbf{m}}\,.
\nn
\end{align}
Its eigenvalue on the eigenvector 
discussed above will be denoted by $Q(\z,\kappa)$.
The main role in TBA is played by the function:
\begin{align}
\frak{a}(\z,\kappa)=\frac{d(\z)Q(\z q,\kappa )}
{a(\z)Q(\z q ^{-1},\kappa)}\,,\label{afrak}
\end{align}
where
$$a(\z)=(1-q\z ^2)^\mathbf{n},\quad d(\z)=(1-q^{-1}\z^2)^\mathbf{n}\,.$$
It follows from the Baxter equation that the solutions to
the equation
$$\frak{a}(\z,\kappa)=-1\,,$$
are the zeros of either $Q(\z,\kappa)$ or $T(\z, \kappa)$. The
function $\frak{a}(\z,\kappa)$ satisfies the nonlinear
integral equation  \cite{KBP,Klumper,DDV}:
\begin{align}
\log \frak{a}(\z,\kappa)=-2\pi i\nu\kappa 
+\log \( 
\frac{d(\z)}
{a(\z)}   \)
-\int\limits _\gamma K(\z/\xi)\log \(1+ \frak{a}(\xi,\kappa)\)
\frac {d\xi^2}{\xi ^2}\,,
\label{DDV}
\end{align}
where the cycle $\gamma$ goes around the zeros of 
$Q(\z,\kappa)$ (Bethe roots)
in the {\it clockwise} direction, 
as opposed to all other contours.

We shall need slightly more general kernel
than $K(\z/\xi)$, so, let us define them together. First, we introduce operations: 
\begin{align}
&\Delta _\z f(\z)=f(\z q)-f(\z q^{-1})\label{deltas}\,,\\
&\delta ^-_\z f(\z)=f(\z q)-\rho(\z|\kappa,\kappa +\al)f(\z)\,.\nn
\end{align}
Then
\begin{align}
K(\z,\al)=\frac 1{2\pi i}\Delta _{\z}\psi (\z,\al) ,\quad
K(\z)=K(\z,0)\,.\label{defK}
\end{align}

We shall use the  the following notation:
$$f\star g= \int\limits _\gamma f(\eta)g(\eta)dm(\eta)\,,$$
where the measure is given by
\begin{align}dm(\eta )=\frac {d\eta ^2}{ \eta ^2\rho (\eta|\kappa, \kappa +\al)
\(1+\frak{a}(\eta,\kappa)\)}\,.\label{measure}
\end{align}

Now we introduce the resolvent of certain integral operator 
\begin{align}
\Rdr
- \Rdr       \star     K_{\al}
=K_{\al}\,,
\label{eqR}
\end{align}
where $K_{\al}$ stands for the integral operator with 
the kernel $K(\z/\xi,\al)$.
Introducing  further the two kernels
\begin{align}
\quad f_\mathrm{left} (\z,\xi)=\textstyle{\frac 1 {2\pi i}\ }\delta ^-_\z\psi(\z/\xi,\al),\quad 
f_\mathrm{right}(\z,\xi)=\delta ^-_\xi \psi (\z/\xi,\al)\,,
\label{flr}
\end{align}
we are ready to write the definition of \cite{BG} cleaning it from irrelevant 
auxiliary objects and taking into account the modification \eqref{omega-tilde}:
\begin{align}
\textstyle{\frac 1 4}\omega (\z,\xi|\kappa,\al)=
&\(f_\mathrm{left}\star   f_\mathrm{right}+   f_\mathrm{left}\star           
\Rdr
\star f_\mathrm{right}\)(\z,\xi)
 \label{defomega}
\\
&
+\delta ^-_\z\delta ^-_\xi\Delta ^{-1}_\z\psi(\z/\xi,\al)
\,.\nn
\end{align}

\section{ Introducing screening operators on the lattice}\label{screenings}

Now we want to describe an important generalisation of 
results of \cite{HGSIII}.
We have three  constants: $\al$ which
defines the tail of the operator and $\kappa$, $\kappa +\al$
which define the twist of Matsubara transfer matrices at
$+\infty$ and $-\infty$. 
As it has been said in the introduction we need more freedom.
Let us explain how an additional parameter $s\in\mathbb{Z}$
can be introduced into $Z^{\kappa}$. We shall see later
that in the scaling limit introduction of this
parameter leads to emancipation of   $\kappa+\al$ from $\kappa$ and $\al$.

Let us  consider the trace
\begin{align}
\Tr _{\mathrm{S}}\Tr _{\mathbf{M}}\Bigl(Y^{(-s)}_{\mathbf{M}}
T_{\mathrm{S},\mathbf{M}}q^{2\kappa S+2(\al -s) S(0)}
\mathcal{O}^{(s)}\Bigr)
\,,\label{trY}
\end{align}
where $Y^{(-s)}_{\mathbf{M}}$ carries spin $-s$. For definiteness we suppose
$s>0$. It follows from the ice condition that in this situation
the operator $\mathcal{O}^{(s)}$ must have spin $s$.

We assume that among the eigenvectors of the transfer matrices
$T_\mathbf{M}(\z,\kappa)$ and $T_\mathbf{M}(\z,\kappa +\al-s)$
there are maximal ones $|\kappa\rangle$, $|\kappa+\al-s,s\rangle$
with eigenvalues $T(\z,\kappa)$  and
$T(\z,\kappa+\al -s,s)$ which are defined by the requirement
that  
$T(1,\kappa)\cdot
T(1,\kappa+\al -s,s)$ is of maximal absolute value among all
the pairs of eigenvectors. We assume further the generality
condition:
\begin{align}
\langle \kappa |Y^{(-s)}_{\mathbf{M}}
|\kappa+\al-s,s\rangle\ne 0  \label{scalarY}
\end{align}
Obviously
the difference between spins of $|\kappa+\al-s,s\rangle$ and
$|\kappa\rangle$ must be equal to $s$. We make the technical assumption that spin of $|\kappa\rangle$ remains equal to zero.

The natural
idea is to create the operators 
$q^{2(\al-s) S(0)}\mathcal{O}^{(s)}$
by $\bb^*$, $\cb^*$, $\tb^*$ having an excess of operators $\bb^*$:
\begin{align*}
q^{2(\al -s)S(0)}\mathcal{O}^{(s)}
&=\bb^*(\xi _1)\cdots \bb ^*(\xi _s)
\nn\\
&\times\bb ^*(\z ^+_m)\cdots \bb ^*(\z ^+_1)
\cb ^*(\z ^-_1)\cdots \cb ^*(\z ^-_m)
\tb^*(\z _1^0)\cdots \tb^*(\z _n^0)
\bigl(q^{2\al S(0)}\bigr)\,.
\end{align*}

However, the formulae
\eqref{mainb} are not applicable in this case. Let us explain why it is so.

The method of \cite{HGSIII} requires to start the consideration
by the operator $\bb ^*(\xi _1)$ which is the closest to $T_{\mathrm{S},
\mathbf{M}}q^{2\kappa S}$. 
The operator $\xi_1^{-\al} \bb _\mathrm{rat}^*(\xi_1)(X)$
is a meromorphic function  of $\xi _1^2$
with singularities at the points $(\z ^-_j )^2$.
It satisfies $\mathbf{n}+1$ normalisation conditions 
\cite{HGSIII}. 
The additional term in \eqref{modtbc} is of the form satisfying \eqref{mainb} from the very
beginning.
These singularities and normalisation conditions are studied algebraically, they
do not change comparing to \cite{HGSIII} where the spin of $X$ was
equal to $-1$. However, the behaviour at zero changes:
in the present case the spin of $X$ equals $s-1$, and according to
\cite{HGSII} $\xi _1^{-\al} \bb ^*(\xi _1)(X)$ does not vanish at zero. Generally,
\begin{align}
\z^{-\al} \bb ^*(\z)(X)=
\sum\limits _{j=0}^{s-1}\z ^{-2j} \bb ^*_{\infty,j}(X)+ 
\z^{-\al} \bb^* _{\mathrm{reg}}(\z)(X),\quad X\in\mathcal{W}_{\al-s+1,  s-1}
\,,
\label{behb}
\end{align}
where $ \z^{-\al} \bb^* _{\mathrm{reg}}(\z)(X)$  vanishes at zero.
But
in
\cite{HGSIII} 
the fact that 
$\z^{-\al} \bb ^*(\z)(X)$
vanishes at zero for $\mathrm{spin}(X)=-1$ was important
when  deriving  \eqref{mainb}.  As a result in the present case
we do not have  enough conditions to
define $\omega $.

Let us turn this problem into advantage. 
The operators $\bb ^*_{\infty,j}$, $j=0,\cdots s-1$
constitute a finite Grassmann algebra. So, we just 
consider $\xi _j\to 0$ and replace 
$\bb^*(\xi _1)\cdots \bb ^*(\xi _s)$ by
$\bb^*_{\infty,s-1}\cdots \bb^*_{\infty,0}$.
Now move one of the remaining operators 
$\bb^*$, namely, $\bb^*(\z^+_1)$ to the left.
Obviously, $(\z^+_1)^{-\al}\bb^*(\z^+_1)$ vanishes as $(\z^+_1)^2\to 0$
because the singular part disappears due to
multiplication by $\bb^*_{\infty,s-1}\cdots \bb^*_{\infty,0}$.
Effectively, this operator reduces to $\bb^* _{\mathrm{reg}}(\z^+_1)$.

It is not hard to see that we can introduce $\omega(\z,\xi)$ and obtain  
\eqref{maint}, \eqref{mainb}, \eqref{mainc} for the functional
on $\mathcal{W}_{\al,0}$: 
\begin{align*}
&Z^{\kappa,s}\Bigl\{q^{2\al S(0)}\mathcal{O}
\Bigr\}
=\frac{\Tr _{\mathrm{S}}\Tr _{\mathbf{M}}\Bigl(Y_\mathbf{M}^{(-s)}T_{\mathrm{S},\mathbf{M}}\ q^{2\kappa S}\mathbf{\bb ^*}_{\infty, s-1}
\cdots\mathbf{\bb^*}_{\infty,0}
\bigl(q^{2\al S(0)}\mathcal{O}\bigr)\Bigr)}
{\Tr _{\mathrm{S}}\Tr _{\mathbf{M}}\Bigl(Y_\mathbf{M}^{(-s)}T_{\mathrm{S},\mathbf{M}}\ q^{2\kappa S} \mathbf{\bb ^*}_{\infty, s-1}
\cdots\ \mathbf{\bb^*}_{\infty,0}
\bigl(q^{2\al S(0)}\bigr)\Bigr)}
\,.
\end{align*}
The function $\rho$ changes in the most natural way to 
\begin{align}
\rho(\z|\kappa,\kappa +\al,s)=\frac{
T(\z,\kappa +\al-s,s)}
{T(\z,\kappa )} \,.\label{rhos}
\end{align}
The function $\omega (\z,\xi|\kappa,\al, s)$ is defined by 
\eqref{defomega}, replacing  $\rho (\z|\kappa,\kappa +\al)$ by
$\rho(\z|\kappa,\kappa +\al,s)$ 
but keeping the same $\frak{a}(\z,\kappa)$.  

Let us
discuss one important  property of Bethe vector of spin $s$.
The basic object in the theory are the transfer matrix
$T_\mathbf{M}(\z,\kappa +\al-s)$ and 
 Baxter's $Q$-operator $Q_\mathbf{M}(\z,\kappa +\al-s)$. 
 Their eigenvalues on the vector $|\kappa +\al-s, s\rangle$
 have the following analytical properties: $T(\z,\kappa +\al-s,s)$
 is a polynomial of $\z ^2$ of degree $\mathbf{n}$, 
 $$Q(\z,\kappa +\al-s,s)=\z ^{-\al -\kappa +2s}A(\z
 ,\kappa +\al-s,s)\,,$$
 where $A(\z ,\kappa +\al-s,s)$ 
is a polynomial in $\z^2$ of degree $\mathbf{n}/2-s$.
 In terms of $T$ and $A$ the Baxter equation reads
 \begin{align}
 &T(\z,\kappa +\al-s,s)
  A(\z 
 ,\kappa +\al-s,s)\label{sbaxter}\\
&=q^{-\kappa '}d(\z) A(\z q
 ,\kappa +\al-s,s)+
q^{\kappa'}a(\z) A(\z q^{-1}
 ,\kappa +\al-s,s)\,.
\nn
 \end{align}
 with
 \begin{align}
 \kappa '=\kappa +\al +2\textstyle{\frac {1-\nu}{\nu}}s\,. \label{kappaprime}
 \end{align}
Formally, $\kappa'$ is defined modulo $2\mathbb{Z}/\nu$. 
However, the eigenvalues are
multi-valued functions of $\kappa '$ and we have to be 
careful about the choice of the branch. 
The choice in \eqref{kappaprime} is consistent 
with the semi-classical limit $\nu\to 1$. 
We shall return to this point when we 
discuss the scaling limit.
 
 Notice that if we write the Baxter equation
 for twist $\kappa'$ and spin $0$:
 \begin{align}
 T(\z,\kappa')
 & A(\z 
 ,\kappa ')
=q^{-\kappa '}d(\z) A(\z q
 ,\kappa')+
q^{\kappa'}a(\z) A(\z q^{-1}
 ,\kappa ')\,,
\label{sbaxter1}
 \end{align}
 it looks exactly the same as \eqref{sbaxter} if we identify
 $T(\z,\kappa')$ with $T(\z,\kappa +\al-s,s)$
 and $A(\z 
 ,\kappa')$ with $A(\z
 ,\kappa +\al-s,s)$. There is,
 however an important difference: the polynomial $A(\z 
 ,\kappa ')$ in \eqref{sbaxter1} is of degree $\mathbf{n}/2$
 while the polynomial $A(\z 
 ,\kappa +\al-s,s)$ 
in \eqref{sbaxter} is of degree $\mathbf{n}/2-s$.
 Still, this similarity will be very important for us later when we shall discuss the scaling limit.
 
 Similarly to the previous discussion we can consider an operator $Y^{(s)}_\mathbf{M}$ which carries positive spin $s$.
 Then the operator $\cb ^*(\z)$ will have nontrivial behaviour at $\z ^2 =0$:
 \begin{align}
\z^{\al} \cb ^*(\z)(X)=
\sum\limits _{j=0}^{s-1}\z ^{-2j} \cb ^*_{\infty,j}(X)+ 
\z^{\al} \cb^* _{\mathrm{reg}}(\z)(X), \quad X\in \mathcal{W}_{\al +s-1   ,-s+1}
\,,
\label{behc}
\end{align}
and we can repeat the entire procedure using $\cb ^*_{\infty,j}$
instead of $\bb ^*_{\infty,j}$. So, $s$ in \eqref{kappaprime} can take any integer value.

Obviously, what we are doing here is nothing else but introducing
screening operators on the lattice. This is important for relating
to the CFT. The screening operators $\bb ^*_{\infty,j}$ anticommute among
themselves, the same is true for $\cb ^*_{\infty,j}$. Naively, one could say
that $\bb ^*_{\infty,j}$ anticommute with $\cb ^*_{\infty,i}$, but this does not
make sense because the product of these  operators do not act nontrivially on any subspace of 
$\mathcal{W}^{(\al)}$.

\begin{rem}\label{descendants}
Let us mention one more, less dramatic, generalisation of the results of 
\cite{HGSIII}.
Clearly the functional $Z^{\kappa,s}$ is independent of 
the choice of $Y^{(s)}_\mathbf{M}$ provided 
the condition \eqref{scalarY} is satisfied.
However different choices are also possible. 
For instance one can take any eigenvector $|A\rangle$
and eigencovector $\langle B|$ of the Matsubara 
transfer matrices and consider the projector $|A\rangle \langle B|$. 
The main formula is applicable in 
this more general setting. 
\end{rem}

\section{Scaling limit}\label{scaling}

We take $\al$, $\kappa$ to be real,  
and restrict our consideration to the region 
\begin{align}
\bigl|\kappa\bigr|<1,\quad \bigl|\kappa '\bigr|<1\,.
\label{domalkappa}
\end{align}
Then we continue analytically.

The crucial data for our construction are the 
eigenvectors  $|\kappa+\al-s,s\rangle$, 
$|\kappa\rangle$. Consider the second of them. 
The corresponding eigenvalue $T(\z,\kappa)$
corresponding to maximal eigenvector $|\kappa\rangle$ of spin zero.
Denote by $Q(\z,\kappa)$ the eigenvalue of 
the $Q$-operator
on $|\kappa\rangle$. 
These eigenvalues have the form 
\begin{align}
&Q(\z,\kappa)=\z^{-\kappa}\prod\limits _{p=1}^{\mathbf{n}/2}
(1-\z ^2/\xi^2_p)\,,\nn
\\
&T(\z,\kappa)=(q^{\kappa}+q^{-\kappa})
\prod\limits _{p=1}^{\mathbf{n}}
(1-\z ^2/\theta ^2_p)\,.\nn
\end{align}
In the domain $|\kappa|<1$, 
the vector $|\kappa\rangle$ is uniquely characterised by the 
two requirements for the roots:  
$\xi ^2_k,\theta ^2_j\in \mathbb{R}$, 
and  
$\xi_k^2>0>\theta^2_j$
for all $k,j$.
Let us study the behaviour of the Bethe
roots $\xi _p$ as $\mathbf{n}\to\infty$. 
Suppose they are numbered 
in the order $\xi_1^2 <\xi_2^2<\xi_3^2<\cdots$. 
In the limit $\mathbf{n}\to\infty$ they are 
subject to the Lieb distribution \cite{Lieb}
: for $1\ll m$ we have
\begin{align}
\frac{\xi _{m+1}}{\xi _m}-1=
\frac{\pi\nu}{\mathbf{n}}\bigl(\xi _m^{\frac 1 \nu}+\xi _m^{-\frac 1 \nu}
\bigr)+
O\Bigl(\frac 1 {\ \mathbf{n}^2}\Bigr)\,.\label{lieb}
\end{align}
We are interested in the 
Bethe roots which are not very far from $\z ^2=0$.
In other words we assume 
$1\ll m \ll\mathbf{n}$. 
Since $\xi_m$ is small, we can drop the 
term $\xi_{m}^{\frac 1 \nu}$ in the \eqref{lieb}, 
obtaining
\begin{align}
\xi _m\simeq \(\pi \frac{m}{\mathbf{n}}\)^{\nu}\,.\label{law}
\end{align}
Similar power law  is obeyed  by $\theta _m$.

So far we have concentrated on the ground states in Matsubara direction. 
But according to Remark \ref{descendants} the main formulae 
can be generalised to arbitrary Bethe states. 
There are 
low-lying excited states 
which satisfy \eqref{law}, and to which the same
analysis as for the ground states apply.  
Readers who are familiar with the papers 
by Bazhanov, Lukyanov and Zamolodchikov 
\cite{BLZI,BLZII,BLZIII} would immediately say that 
in the limit we obtain the CFT 
transfer matrices treated by them. 
We shall come to this relation later in section \ref{reviewBLZ}.

The formulae \eqref{maint}, \eqref{mainb} 
and \eqref{mainc} imply the explicit expression
\begin{align}
&Z^{\kappa,s}\bigl\{\tb^*(\z^0_1)\cdots \tb^*(\z^0_p)
\bb^*(\z^+_1)\cdots \bb^*(\z^+_r)
\cb^*(\z^-_r)\cdots \cb^*(\z^-_1)\bigl(q^{2\al S(0)}\bigr)\bigr\}
\label{main}
\\
&=\prod\limits _{i=1}^p 2\rho (\z _i^{0}|\kappa,\kappa+\al,s)\times
\det \left(\omega(\z^+_i,\z ^-_j|\kappa,\al,s)
\right)_{i,j=1,\cdots, r}\,.\nn
\end{align}

We consider the scaling limit in the Matsubara direction, 
\begin{align}
\mathbf{n}\to \infty, \quad a\to 0,
\quad \mathbf{n}a=2\pi  R\ \ \mathrm{fixed}\,.
\label{scalinglimit} 
\end{align}
 
Let us
introduce the following strangely looking notation
\begin{align}
\bar{a}=Ca\,,\label{abar}
\end{align}
where $C$ is some $\nu$-dependent constant which will be needed
for fine tuning comparing the scaling limit with CFT.

The following limits exist for finite $\la$:
\begin{align}
&\Tb(\la,\kappa)= 
\lim_{\mathbf{n}\to\infty,\ a\to 0,\ 
2\pi  R=\mathbf{n}a} T(\la \bar{a}^\nu,\kappa)\,,
\label{QQQQ}
\\
&\Qb(\la ,\kappa)
=\lim_{\mathbf{n}\to\infty,\ a\to 0,\ 
2\pi   R=
\mathbf{n}a} \bar{a}^{\nu\kappa}Q(\la \bar{a} ^\nu,\kappa)\,.
\nn
\end{align}
The eigenvalues of 
$\Tb(\la,\kappa)$, 
$\Qb(\la ,\kappa)$ 
are given by convergent infinite products
due to \eqref{law}. In particular, it is easy to see from \eqref{law}
that the following asymptotics hold:
\begin{align}
\log \Qb(\la,\kappa)
\ \ \ \sim \hskip -.8cm{}_{{}_{\la ^2\to-\infty}}\  2\pi R
\cdot \frac {C}{\sin \frac {\pi}{2\nu}}\ (-\la ^2)^{\frac 1 {2\nu}}\,.
\label{asymDsc}
\end{align}
Certainly, the limits exist for eigenvalues corresponding to any
eigenvectors satisfying \eqref{law}. 

Now we turn to the operators  
$T_\mathbf{M}(\z,\kappa+\al-s)$,
$Q_\mathbf{M}(\z,\kappa+\al-s)$ for which the eigenvectors
of spin $s$ are considered. 
The definitions of 
$\Tb(\la ,\kappa+\al-s,s)$, 
$\Qb(\la ,\kappa +\al-s,s)$ are the same as before.
The important statement is that
\begin{align}
&\Tb(\la ,\kappa+\al-s,s)=\Tb(\la ,\kappa')\,,
\quad \ 
\Qb(\la ,\kappa +\al-s,s)=
\Qb(\la ,\kappa')\,,\nn
\end{align}
where $\kappa '$ is given by \eqref{kappaprime}. 
The right hand sides are understood as "correct" analytical
continuations from spin $0$ sector. We have to explain two points:
first, what is the reason that in the scaling limit the eigenvalues in the spin $s$ sector
equals analytical continuations of those in the spin $0$ sector; second,
what  we mean by ``correct" analytic continuations.
 The first point is simple: recall
the discussion concerning the similarity
of the equations \eqref{sbaxter}, 
\eqref{sbaxter1}. The only difference between them was the number
of Bethe roots, but this number is infinite in the scaling limit, so,
the difference disappears. On the other hand the eigenvalue
$\Tb(\la ,\kappa)$
is a  multi-valued function
of $\kappa$, so we have to explain the choice of its branch.
At this point we refer to the semi-classical domain : $\nu$
close to $1$. We take a good branch at this domain, and then continue analytically. Notice that
$T(0,\kappa)=2\cos (\pi\nu\kappa)$. We require that introducing $s$
does not deviate us far from this value. This was the reason for
choosing the definition \eqref{T0} because with this definition we 
have 
$$
T(0,\kappa+\al-s,s)=2\cos (\pi\nu(\kappa+\al
+2\textstyle{\frac {1-\nu}{\nu}} s))\,,
$$
which stays close to $2\cos (\pi\nu\kappa)$ for all $s$ if $\nu$ is close to $1$.

From now on we shall often consider $\kappa$ and $\kappa '$ as arbitrary numbers implying the
possibility of analytical continuation from values \eqref{kappaprime}.

Using $\Qb(\la,\kappa)$, $\Qb(\la,\kappa')$ 
we obtain finite limits
\begin{align}
&\rhob(\la|\kappa,\kappa ')
=\ \lim_{\mathbf{n}\to\infty,\ a\to 0,\ 
2\pi   R=\mathbf{n}a}\ \rho(\la \bar{a}^{\nu}|\kappa,\al,s),
\label{limro}\\
&4\ \omegab(\la,\mu|\kappa,\kappa ',\al)
=  \lim_{\mathbf{n}\to
\infty,\ a\to 0,\ 2\pi   R=\mathbf{n}a}
\ \omega(\la \bar{a} ^\nu,\mu \bar{a}^\nu|\kappa,\al,s)\,,
\label{limro2}
\end{align}
where $\kappa '$ is given by \eqref{kappaprime}, and then for 
$\rhob(\la|\kappa,\kappa ')$,
$\omegab(\la,\mu|\kappa,\kappa ',\al)$ 
the analytical continuation with respect to $\kappa '$ is used.

According to Remark  \ref{descendants}
 \eqref{main} remains valid
for  any Bethe states in Matsubara direction. 
It has been already said that to Bethe
states close to the ground states the scaling
procedure applies.
We shall argue later on that 
these vectors span the Verma module of chiral CFT. 
So, we would like to use the right hand side
of \eqref{main} in order to consider the scaling limit in the space direction 
\begin{align}
j a=  x\ \ \mathrm{finite}. \nn
\end{align}
In our setting it amounts to considering the operators:
\begin{align}
&2\taub ^*(\la)=
\lim_{a\to 0}\tb ^*(\la \bar{a}^\nu), 
\label{deftbg}
\\
&2\betab ^*(\la)
=\lim_{a\to 0}\bb ^*(\la \bar{a}^\nu)  ,\quad
2\gammab ^*(\la)=\lim_{a\to 0}\cb ^*(\la \bar{a}^\nu)\,,
\nn
\end{align}
and 
\begin{align*}
\Phi_{\al}(0)=\lim_{a\to 0} q^{2\al S(0)}, 
\end{align*}
so that \eqref{main} gives in the scaling limit
\begin{align}
&Z_R^{\kappa,\kappa '}\bigl\{\taub^*(\la^0_1)\cdots \taub^*(\la^0_p)
\betab^*(\la^+_1)\cdots \betab^*(\la^+_r)
\gammab^*(\la^-_r)\cdots \gammab^*(\la^-_1)
\bigl(\Phi_\al(0)\bigr)
\bigr\}
\label{main2}
\\
&=\prod\limits _{i=1}^p 
\rhob(\la _i^{0}|\kappa,\kappa ')
\times
\det 
\left(
\omegab(\la^+_i,\la ^-_j|\kappa,\kappa ',\al)
\right)_{i,j=1,\cdots, r}\,.
\nn
\end{align}

We understand the formula \eqref{main} as giving the 
expectation values of certain non-local operators making
contact with quasi-local ones near $\z ^2=1$.
After introducing $a$ this point moves to $\la ^2=\bar{a}^{-2\nu}$, 
and in the scaling limit it goes further to $\la ^2=+\infty$. 
It should not be a surprise that this limit is described by CFT. 
To make this statement precise, we have to establish certain 
asymptotic properties of 
$\rhob(\la|\kappa,\kappa ' )$, 
$\omegab(\la,\mu|\kappa,\kappa ' ,\al)$ 
for $\la ^2,\mu^2\to+\infty$.
But first we shall need some information about 
the integrable structure of CFT.

\section{CFT on a cylinder and three point functions}\label{CFT}

In this section we introduce our notation  
concerning CFT, and collect 
a few facts which will be relevant to the subsequent sections.

Consider chiral CFT on a cylinder 
$\cyl=\mathbb C/2\pi i R\mathbb Z$ 
with circumference $2\pi R$, the points 
$x$ and $x+2\pi i R$ being identified. 
Along with the local coordinate $x$, 
we shall also use the global coordinate 
\begin{align*}
z=e^{-\frac xR}\,. 
\end{align*}
The two points $x=-\infty,\infty$ on the boundary of $\cyl$ 
correspond respectively to the points $z=\infty,0$ 
on the Riemann sphere. 

Let
\begin{align*}
T(x)=\sum_{n=-\infty}^\infty l_nx^{-n-2}
\end{align*}
be the energy-momentum tensor in the coordinate $x$,   
where the $l_n$'s satisfy the commutation 
relations of the Virasoro algebra with the central charge
\begin{align}
&c=1-\frac{6\,\nu^2}{1-\nu}\,.
\label{central}
\end{align}
The energy-momentum tensor in the coordinate $z$, 
\begin{align*}
\tilde T(z)=\sum_{n=-\infty}^\infty L_nz^{-n-2},
\end{align*}
is related to $T(x)$ via the transformation rule 
$\tilde T(z)(dz)^2=(T(x)-(c/12)\{z;x\})(dx)^2$. 
Here $\{z;x\}$ denotes the Schwarzian derivative. 
In turn, $T(x)$ is written as 
\begin{align*}
T(x)=\frac1{R^2}\Bigl(\sum_{n=-\infty}^\infty 
L_ne^{\frac{nx}R}-\frac c{24}\Bigr)\,.
\end{align*}
The Virasoro algebra acts on a local field
$O(y)$ by the contour integral
\begin{align}
({\bf l}_nO)(y)=\int\limits_{C_y}\frac{dx}{2\pi i}
(x-y)^{n+1}T(x)O(y)\,,
\end{align}
where ${C_y}$ encircles the point $y$ anticlockwise.

From now on, we fix a primary field 
$\phi_\Delta (y)$ with the scaling dimension $\Delta$: 
$$
(\mathbf{l}_0\phi_\Delta)(y)=\Delta\phi_\Delta(y),\quad 
(\mathbf{l}_n\phi_\Delta)(y)=0\quad (n>0)\,, 
$$
and study the expectation values
\begin{align}
\langle 
T(x_k)\cdots T(x_1) \phi_\Delta(y) \rangle_{\Delta_+,\Delta_-} \,.
\label{3pt}
\end{align}
The suffix indicates that 
we consider \eqref{3pt} in the presence of  
two other primary fields inserted at $x=\pm\infty$.
More precisely, we impose the boundary conditions 
\begin{align}
&\lim\limits_{x\rightarrow\pm \infty}T(x)
=\frac1{R^2}\left(\Delta_\pm-\frac c{24}\right)\,
\label{bdry-cond}
\end{align}
inside the expectation values \eqref{3pt},  
where  $\Delta_\pm$ are the conformal dimensions of the 
inserted primary fields.     
For readers who prefer the language of representation theory, 
we are considering a highest weight vector $|\Delta_+\rangle$
at $z=0$ satisfying 
$L_n|\Delta_+\rangle=\delta_{n,0}\Delta_+|\Delta_+\rangle$ 
($n\geq0$),
and a co-vector $\langle\Delta_{-}|$ at $z=\infty$
satisfying
$\langle\Delta_{-}|L_n=\delta_{n,0}\Delta_{-}\langle\Delta_{-}|$ 
($n\leq0$).

The singular part of \eqref{3pt} is known from OPEs. 
In order to write them in the coordinate $x$, 
it is useful to introduce the function 
\begin{align*}
\chi(x)=\frac12\coth\left(\frac x{2R}\right)
=\sum\limits_{n=0}^\infty\frac {B_{2n}}{(2n)!}
\left(\frac{x}{R}\right)^{2n-1}\,.
\end{align*}
Here $B_0=1,B_2=1/6,B_4=-1/30,\cdots$ are the Bernoulli numbers. 
The main OPEs then read, as $x\to y$,  
\begin{align}
T(x)T(y)&=
-\frac{c}{12 R}\chi'''(x-y)
-\frac{2T(y)}{R}\chi'(x-y)
+\frac{T'(y)}{R}\chi(x-y)+O(1)\,,
\label{OPE1}\\
T(x)\phi_\Delta(y)&=
-\frac{\Delta \phi_\Delta(y)}{R}\chi'(x-y)
+\frac{\phi_\Delta'(y)}{R}\chi(x-y)+O(1)\,,
\label{OPE2}
\end{align}
where the prime stands for the derivative. 

The OPEs \eqref{OPE1}, \eqref{OPE2}, 
combined with \eqref{bdry-cond} and the behaviour  
\begin{align*}
\chi(x)=\pm\frac12+O(e^{\mp x/R})
\quad (x\rightarrow\pm\infty), 
\end{align*}
allow us to write the conformal Ward-Takahashi 
identity which determines \eqref{3pt} recursively:
\begin{align*}
&\langle T(x_k)\cdots T(x_1) \phi_\Delta(y) \rangle_{\Delta_+,\Delta_-}
\\
&\quad
=-\frac{c}{12R}\sum_{j=2}^k\chi'''(x_1-x_j)
\langle T(x_k)\cdots \overset{j}{\widehat{\phantom{T}}}\cdots
T(x_2) \phi_\Delta(y) \rangle_{\Delta_+,\Delta_-}
\\
&\quad+
\Bigl\{\sum_{j=2}^k\bigl(-\frac{2}{R}\chi'(x_1-x_j)
+\frac{1}{R}(\chi(x_1-x_j)
-\chi(x_1-y))\frac{\partial}{\partial x_j}\bigr)
-\frac{\Delta}{R}\chi'(x_1-y)
\\
&\quad 
+(\Delta_+-\Delta_-)\frac{1}{R^2}\chi(x_1-y)
+\frac{1}{2R^2}(\Delta_++\Delta_-)-\frac{c}{24R^2}\Bigr\}
\langle T(x_k)\cdots T(x_2) \phi_\Delta(y) \rangle_{\Delta_+,\Delta_-}\,.
\end{align*} 
From these one can extract, for example, 
\begin{align*}
\frac{\langle (\mathbf{l}_{-n}\phi_\Delta)(y)\rangle_{\Delta_+,\Delta_-}}
{\langle\phi_\Delta(y)\rangle_{\Delta_+,\Delta_-}}
=
\begin{cases}
\displaystyle{
\frac{\delta_{n,2}}{2R^2}
({\scriptstyle\Delta_++\Delta_{-}}-\frac c{12})
-\frac{B_n\Delta}{n(n-2)!R^n}}&\quad(n:\hbox{even});\\[7pt]
\displaystyle{\frac{(\Delta_+-\Delta_{-})B_{n-1}}{(n-1)!R^n}}
&\quad(n:\hbox{odd}).\\
\end{cases}
\end{align*}
In general, the normalised three point function 
of any particular descendant 
\begin{align}
\frac{\langle 
\bigl(
\mathbf{l}_{-n_k}\cdots\mathbf{l}_{-n_1}
\phi_\Delta\bigr)(y)
\rangle_{\Delta_+,\Delta_-}}
{\langle \phi_\Delta(y)\rangle_{\Delta_+,\Delta_-}}
\label{3pt2}
\end{align}
is determined as a polynomial in $\Delta,\Delta_+,\Delta_-$. 

In writing these formulas, we have tacitly assumed that 
$\langle \phi_\Delta(y)\rangle_{\Delta_+,\Delta_-}$ 
is non-trivial. 
Actually, in the theory with $c<1$, 
for a given generic value of $\Delta$   
there is a discrete but an infinite collection 
of such $\Delta_+,\Delta_-$. 
In view of the polynomial dependence mentioned above, 
one can regard \eqref{3pt2} as a linear functional, 
defined for arbitrary  $\Delta,\Delta_+,\Delta_-$, 
on the Verma module 
consisting of the descendants of $\phi_\Delta(y)$. 
Henceforth we shall adopt this point of view. 
In later sections we shall use the parametrisation
$\Delta=\Delta_\al$, 
$\Delta_+=\Delta_{\kappa+1}$, 
$\Delta_-=\Delta_{-\kappa'+1}$, 
and 
write this functional as 
\begin{align}
Z^{\kappa,\kappa'}_R
\Bigl\{X(y)\Bigr\}=
\frac{\langle X(y) 
\rangle_{\Delta_{\kappa+1},\Delta_{-\kappa'+1}}}
{\langle \phi_{\Delta_\al}(y)
\rangle_{\Delta_{\kappa+1},\Delta_{-\kappa'+1}}}
\label{ZR}
\end{align}
where $X(y)$ is a descendant of $ \phi_{\Delta_\al}(y)$.

In \cite{Zam}, A. Zamolodchikov   
introduced the local integrals of motion which survive
the $\phi _{1,3}$-perturbation of CFT. 
They act on local operators as
\begin{align}
(\mathbf{i}_{2n-1}O)(y)
=\int _{{C_y}}\frac {d x}{2\pi i}h_{2n}(x)O (y)\,
\quad (n\ge 1).
\label{i}
\end{align}
The densities $h_{2n}(y)$ are certain
descendants of the identity operator $I$. 
The simplest examples are 
\begin{align}
h_{2}(x)=(\mathbf{l}_{-2}I)(x)=T(x),\quad 
h_{4}(x)=(\mathbf{l}_{-2}T)(x)\,,
\end{align}
for which we have
\begin{align*}
&(\mathbf{i}_{1}O)(y)=(\mathbf{l}_{-1}O)(y),\quad
(\mathbf{i}_{3}O)(y)=
2\sum_{n=0}^{\infty}(\mathbf{l}_{-n-2}\mathbf{l}_{n-1}O)(y)\,.
\end{align*}

In general, for a descendant $h(x)$ of the identity 
operator, the three point function 
\begin{align*}
\int\limits_{C_y}\frac{dx}{2\pi i}
\langle h(x)O(y)\rangle_{\Delta_+,\Delta_-}
\end{align*}
is reducible to that of $O(y)$. 
Indeed, it can be rewritten as 
\begin{align*}
-\int\limits_{C(u)}\frac{dx}{2\pi i}
\langle h(x)O(y)\rangle_{\Delta_+,\Delta_-}
+\int\limits_{C(v)}\frac{dx}{2\pi i}
\langle O(y)h(x)\rangle_{\Delta_+,\Delta_-},
\end{align*}
where $C(u)$ $(u\in\mathbb R)$ is a circle 
with the real part $u$, starting from $u-\pi Ri$ 
and ending at $u+\pi Ri$. 
Choosing $u\ll {\rm Re}\ y \ll v$ 
and using the boundary conditions at $x\rightarrow\pm\infty$,
one can show that each of 
them reduce to a constant. 

From the above remark it follows that
\begin{align}
\langle \mathbf{i}_{2n-1}\bigl(O(y)\bigr)
\rangle_{\Delta_+,\Delta_-}
=
(I^+_{2n-1}-I^-_{2n-1})\cdot 
\langle O(y) \rangle_{\Delta_+,\Delta_-}\,,
\label{Int-m}
\end{align}
where $I^{\pm}_{2n-1}$ denote the vacuum 
eigenvalues of the 
local integrals of motion on the Verma module 
with conformal dimension $\Delta_\pm$. 
Their explicit formulas for small $n$ can be found in 
\cite{BLZI} (see also section \ref{asymp-loga},  
\eqref{I1}--\eqref{I5} below). 
Notice that, in the special case $\Delta_+=\Delta_{-}$, 
the three point function 
vanishes on the image of the local integrals. 

As mentioned in the introduction, 
we accept the conjectural 
statement that the Verma module is spanned 
by the elements 
$$
\mathbf{i}_{2k_1-1}\cdots \mathbf{i}_{2k_p-1}
\mathbf{l}_{-2m_1}\cdots \mathbf{l}_{-2m_q}(\phi _{\al}(0))\,.
$$
Formula \eqref{Int-m} tells that for 
the computation of the linear functional \eqref{3pt2} 
it suffices to consider the descendants by 
the even Virasoro generators $\{{\bf l}_{-2n}\}_{n\geq1}$.

Before closing this section, let us comment on a point which could be 
a source of confusion. 
The local integrals of motion arise in two different ways.    
At the boundary of the cylinder, they appear as the 
operators $I_{2n-1}$ constructed from the 
modes $L_n$ of the energy-momentum tensor in the coordinate $e^{-x/R}$.   
These operators preserve the subspace of the Verma module of a given degree
and can be diagonalised. 
In the classical limit,  the eigenvalues of $I_{2n-1}$ correspond to the 
values of the integrals of motion 
on quasi-periodic solutions to the KdV hierarchy. 
In contrast, the action of the integrals of motion 
on local fields $\mathbf{i}_{2n-1}$ are 
constructed from the modes $\mathbf{l}_n$ in the coordinate $x$. 
Unlike in the first case, they do not commute with 
$\mathbf{l}_0$ (for example, the first integral of motion is 
$\mathbf{l}_{-1}$). 
They act as a creation part of the Heisenberg algebra.  
In the classical limit, they correspond to the action 
of the Hamiltonian vector 
fields generated by the local integrals of motion.
So it does not make sense to talk about their diagonalisation.

\section{Brief review of BLZ}\label{reviewBLZ}

In a series of papers \cite{BLZI,BLZII,BLZIII}
Bazhanov, Lukyanov and Zamolodchikov (BLZ) 
studied the integrable structure of the chiral CFT on a circle. 
We shall recall these results briefly,   
since they are quite relevant to us.   

It is convenient to write 
the energy-momentum tensor in terms of the chiral boson 
$$
\varphi(x)=
iQ+P\frac{x}{R}
+\sum\limits_{n\ne 0}\frac {a_{n}}n e^{\frac{nx}R}\,.
$$
The operators $P,Q$ are canonically conjugate, 
and the $a_n$'s satisfy the Heisenberg algebra
$$
[P,Q]=\frac{i(1-\nu)}2,
\quad [a_m,a_n]=\frac{m(1-\nu)} 2\delta_ {m+n,0}\,.
$$
The energy-momentum tensor is expressed as 
$$
(1-\nu)T(x)=:\varphi '(x)^2:
+\nu \varphi ''(x)-\frac {1-\nu}{24 R^2}\,.
$$
The chiral vertex operator
\begin{align}
\phi_\al(x)=e^{-\frac{\nu^2\al^2}{4(1-\nu)}\frac{x}{R}}
:e^{\frac{\nu}{1-\nu}\al \varphi(x)}:
\label{primary}
\end{align}
is a primary field of scaling dimension
\begin{align}
\Delta _\al=\frac{\al(\al -2)\nu^2}{4(1-\nu)}.
\label{del-a}
\end{align}
We note that the parameter $\beta^2$ used in \cite{BLZII} 
is identified as 
$$
\beta^2=1-\nu, 
$$  
hence their $q=e^{\pi i \beta^2}$ is our $-q^{-1}$.  
We have also changed the sign of $P, a_n$ and 
the normal ordering convention in \cite{BLZII} to 
\begin{align*}
:e^{\lambda\varphi(x)}:=
e^{\lambda\sum_{n<0}\frac{a_n}{n}e^{nx/R}}
e^{i\lambda Q}e^{\lambda P x/R}
e^{\lambda\sum_{n>0}\frac{a_n}{n}e^{nx/R}}\,,
\end{align*}
which results in the appearance of a scalar factor
$e^{-\frac{\nu^2a^2}{4(1-\nu)}\frac{x}{R}}$
in \eqref{primary}.

The main object studied by BLZ is the universal 
monodromy matrix in CFT. 
It is an element of $U_q(\mathfrak{b}^+)\otimes A_\mathbf{H}$, 
where $U_q(\mathfrak{b}^+)$ is the Borel subalgebra 
of $U_q(\slth)$ generated by $e_0,e_1, h_1$, and 
$A_\mathbf{H}$ is the 
algebra generated by $P,Q, a_{n}$ 
(the suffix $\mathbf{H}$ stands for Heisenberg).
Set 
\begin{align}
&\mathcal{K}_{\mathbf{H}}(x)
=e_0\otimes V_+(x)+e_1\otimes V_-(x)\,,
\quad 
\mathcal{H}_{\mathbf{H}}=-\frac{2P}{1-\nu}\,, 
\label{lop}
\\
&V_\pm(x)
=e^{-(1-\nu)\frac{x}{R}}:e^{\pm 2\varphi(x)}:\,, 
\label{VO}
\end{align}
so that 
$[\mathcal{H}_{\mathbf{H}},V_{\pm}(x)]=\pm2 V_{\pm}(x)$. 
The universal monodromy matrix is defined to be 
\begin{align}
\mathcal{T}_{\mathbf{H}}(\la)=
 \mathcal{P}\exp\( \la\int\limits_{0}^{2\pi R}
\mathcal{K}_{\mathbf{H}}(-iy)dy\)
q^{-\frac{1}{2} h_1\otimes \mathcal{H}_{\mathbf{H}}}\,.
\label{monod}
\end{align} 
Here $\mathcal{P}\exp$ stands for the path ordered exponential.  
Formula \eqref{monod} is understood as a power series 
in $\la$. The integrals in each term 
converge in the domain $1/2<\nu<1$.  
Otherwise divergences occur and a regularisation is needed.

We have considered two maps: $U_q(\frak{b}_+)\to \mathrm{End}(V_a)$ with two-dimensional $V_a$,
and $U_q(\frak{b}_+)\to Osc_A$ with $Osc_A$ being the $q$-oscillator algebra
(see \cite{HGSII} for the notation). 
The images of $\mathcal{T}_{\mathbf{H}}(\la)$ 
under these maps are
denoted by $\mathcal{T}_{a,\mathbf{H}}(\la)$ 
and $\mathcal{T}_{A,\mathbf{H}}(\la)$. 
Then following \cite{BLZII} we define
\begin{align}
&T^{\text{CFT}}_\mathbf{H}(\la)=\Tr _a \(\mathcal{T}_{a,\mathbf{H}}(\la)
e^{-2\pi i (\sigma ^3_a\otimes P)}\)\label{transfer}\,,\\
&Q^{\text{CFT}}_\mathbf{H}(\la)=\la ^{\frac {2P}\nu}
(1-e^{2\pi i P})\Tr _A \(\mathcal{T}_{A,\mathbf{H}}(\la)e^{2\pi i (D_A\otimes P)}\)\,.\nn
\end{align}
There is a slight difference with \cite{BLZII} due to different notation for the
$q$-oscillator algebra.

These operators satisfy the Baxter equation
\begin{align}
T^{\mathrm{CFT}}_\mathbf{H}(\la)Q_\mathbf{H}^{\mathrm{CFT}}(\la)
=Q^{\mathrm{CFT}}_\mathbf{H}(\la q^{-1})
+Q_\mathbf{H}^{\mathrm{CFT}}(\la q)\,.
\label{baxterblz}
\end{align}

An important property of these  
transfer matrices is that they commute with
the local integrals of motion. 
The first local integral of motion 
is nothing but $L_0-\frac {c}{24}$,    
which commutes with the transfer matrices as 
mentioned above.  
Hence each of their eigenstate on a Verma module 
belongs to the subspace of a definite degree. 
In particular, the highest weight vector of 
the Verma module (primary field) is an eigenvector. 

Actually, the local integrals of motion 
are all encoded in the transfer matrix 
$T^{\mathrm{CFT}}_\mathbf{H}(\la)$. The latter 
is known \cite{BLZI} to be 
an entire function of $\la^2$. 
One of the main statements of \cite{BLZI}
is that it has the following asymptotics for $\la^2\to\infty$, 
$\la^2\notin \mathbb{R}_{<0}$, 
\begin{align}
\log(T^{\mathrm{CFT}}_\mathbf{H}(\la))
\sim RC_0\ \la ^{\frac 1 \nu}+\sum\limits_{n=1}^{\infty}
C_n
\la ^{-\frac {2n-1}{\nu}}I_{2n-1}\,,\label{asymptT}
\end{align} 
where 
$C_{n}$ 
are known constants 
which can be found in \cite{BLZII}, they can be also extracted from section \ref{asymp-loga}
of the present paper.
For the moment the only point relevant to us is the fact that 
$$
C_1<0\,. 
$$
This means that for sufficiently large $\la^2$ 
the highest weight vector is the 
eigenvector of $T^{\mathrm{CFT}}_\mathbf{H}(\la)$ 
with the maximal absolute value.
The eigenvalue of $L_0$ on the highest weight vector equals
\begin{align}
\frac 1 {1-\nu}\( {P^2}-\frac{\nu ^2}{4}\)\,.
\label{dimP}
\end{align}

One may wonder why the asymptotic expansion 
\eqref{asymptT} is given as a series in 
the fractional power $\la^{\frac 1\nu}$. 
As explained in \cite{BLZI},    
the reason is that $\la^{\frac 1 \nu}$ is 
a dimensionful quantity having the dimension of 
the inverse length.  
Indeed, consider the $L$-operator \eqref{lop}. 
It must have the dimension
of the inverse length in order that
the exponential in \eqref{monod} be dimensionless. 
But the operators $V_{\pm}(x)$ carry
the anomalous dimension $1-\nu$. 
So, clearly the dimension of $\la$ equals $\nu$.

We shall be interested in the Bethe roots, which are the 
zeros $\la^2=\la^2_n$ of $Q^{\mathrm{CFT}}_\mathbf{H}(\la)$. 
They behave as
$$
\la^2_n=O(n^{2\nu}),\quad n\to\infty\,.
$$
Of equal significance are the zeros 
$\la^2=\mu^2_n$ of $T^{\mathrm{CFT}}_\mathbf{H}(\la)$.
The eigenvalue corresponding to the primary
field has the characteristic property 
\cite{BLZII} that all 
$\la^2_i>0>\mu^2_j$
for all $i,j$.

\section{Conformal limit 
in the Matsubara direction}
\label{contmatsubara}

Let us return to the XXZ model. 
Using the notation introduced in 
section \ref{scaling}, we write the Baxter
equation 
\begin{align}
T_\mathbf{M}(\la \bar{a}^{\nu},\kappa)Q_\mathbf{M}(\la \bar{a}^{\nu},\kappa)
=a(\la \bar{a}^{\nu})Q_\mathbf{M}(\la \bar{a}^{\nu} q^{-1},\kappa)
+d(\la \bar{a}^{\nu})Q_\mathbf{M}( \la \bar{a}^{\nu} q,\kappa)
\,,
\label{baxterXXZ}
\end{align}
where 
$$
a(\la \bar{a}^{\nu})=(1-q \bar{a}^{2\nu}\la^2)^{\mathbf{n}},\quad 
d(\la \bar{a}^{\nu})=(1-q^{-1} \bar{a}^{2\nu}\la^2)^{\mathbf{n}}\,.
$$
We are interested in the maximal eigenvector 
$|\kappa\rangle$ of $T_\mathbf{M}(\la \bar{a}^{\nu},\kappa)$.

We want to consider the limit $\mathbf{n}\to \infty$, $\bar{a}\to 0$, 
while keeping $\mathbf{n} \bar{a}=2\pi   R\cdot C$ and $\lambda$ fixed. 
In this limit, for $1/2<\nu<1$, 
$$
a(\la)\to 1,\quad d(\la)\to 1\,,
$$
so, if we identify
\begin{align}
\nu\kappa
=-2P
\,.\label{nullmode}
\end{align}
the Baxter equation \eqref{baxterXXZ} 
turns into \eqref{baxterblz}.

Now we want to fix the constant $C$ in order to make the
equivalence between 
$\Qb$ 
and $Q^{\mathrm{CFT}}$ exact.
We had the asymptotics \eqref{asymDsc}. On the other hand,
it is known \cite{BLZII} that 
$$\log Q^{\mathrm{CFT}}(\la,\kappa)\sim 
 R\cdot \frac1 {\sqrt{\pi}} \Gamma\(\frac {1-\nu}{2\nu}\)\Gamma \(1-\frac 1{2\nu}\)\Gamma (\nu)^{\frac 1\nu}\ (-\la ^2)^{\frac 1 {2\nu}}\,.$$
So, comparing we see that the agreement is exact if
\begin{align}
C=\  \frac{\Gamma\(\frac {1-\nu}{2\nu}\)}{2\sqrt{\pi}\ \Gamma \(\frac 1{2\nu}\)}\Gamma (\nu)^{\frac 1\nu}\,.
\label{C}
\end{align}

So, we come to 
\begin{align}
\Qb(\la ,\kappa)
=\left.Q^\mathrm{CFT}(\la)\right|_{P=-\frac {\nu\kappa}2}\,.
\end{align}
Let us argue that the vector $|\kappa\rangle$ goes to
the primary field  with the dimension
\begin{align*}
\Delta _{\kappa +1}
=\frac {\nu^2}{4(1-\nu)}(\kappa ^2-1)
\,.
\end{align*}
On the lattice 
the maximal eigenvector $|\kappa\rangle$ is defined 
by the requirement that the eigenvalue 
$T(1, \kappa)$ is of maximal 
absolute value. 
In the scaling limit   
this corresponds to the requirement 
that 
$T^\mathrm{CFT}_\mathbf{H}(\la,\kappa)$
is maximal for $\lambda^2$ large and positive. 
But the asymptotic behaviour of the BLZ 
transfer matrix is given by \eqref{asymptT},
so in the domain 
\begin{align}
-\frac{\pi\nu}{2}<\arg{\la}<\frac{\pi\nu}{2}
\label{argla}
\end{align}
the maximal eigenvalue corresponds to the primary field.
Comparing with \eqref{del-a} and  \eqref{dimP}, 
we find that in the picture of section \ref{CFT} we have 
$$\Delta_{+}=\Delta _{\kappa +1}\,.$$ 
Hence the boundary conditions at $+\infty$ is 
described by $\phi _{\kappa +1}(+\infty)$. 
Similarly, considering the left Matsubara transfer matrix
we find that its 
ground state corresponds to the scaling  dimension
\begin{align}
\Delta_{-}
=\Delta _{-\kappa ' +1}\,.
\label{diml}
\end{align}
In other words, 
in the picture of section \ref{CFT} we have 
the left boundary condition described by
the primary field $\phi_{-\kappa '+1}(-\infty)$.

The argument above is not rigourous, 
because it involves two limits 
which are {\it a priori} non-commutative. 
Nevertheless we believe it makes sense 
because of integrability,   
which stipulates that $|\kappa\rangle$ 
is an eigenvector of $T_\mathbf{M}(\la \bar{a}^{\nu},\kappa)$ 
for all values of $\la$. 
As a supporting argument, we quote 
from \cite{baxter} 
a knowledge that the eigenvalue $T(\z,\kappa)$ 
of the lattice transfer matrix for $\tau =q^{1/2}$ has 
maximal absolute value in the range $|\z|=1$, 
$-\pi\nu/2<\arg \z<\pi\nu/2$.  
This agrees exactly with  \eqref{argla}.

We call the previous reasoning a macroscopic one. 
For completeness let us give a less formal,
microscopic derivation
providing at the same time a constant which is important for physics.
Consider the Matsubara transfer matrix 
$T_\mathbf{M}(\la \bar{a}^\nu,\kappa)$, which is given explicitly by 
\begin{align}
T_\mathbf{M}(\la \bar{a}^{\nu},\kappa)=
\Tr _j \(L_{j,\mathbf{n}}(\la \bar{a}^{\nu})\cdots L_{j,\mathbf{1}}(\la \bar{a}^{\nu})q^{\kappa\sigma ^3_j}\)\,,
\label{TM}
\end{align}
where
$$
L_{j,\mathbf{m}}(\la \bar{a}^\nu)=
\begin{pmatrix}
q^{-\frac 1 2 \sigma ^3_\mathbf{m}}-\bar{a}^{2\nu}\la ^2q^{\frac 1 2 \sigma ^3_\mathbf{m}}&-(q-q^{-1})\la \bar{a} ^{\nu}\sigma ^-_\mathbf{m}\\
-(q-q^{-1})\la \bar{a} ^{\nu}\sigma ^+_\mathbf{m} &
q^{\frac 1 2 \sigma ^3_\mathbf{m}}-\bar{a}^{2\nu}\la ^2q^{-\frac 1 2 \sigma ^3_\mathbf{m}}
\end{pmatrix}_j\,.
$$
Let us make the gauge transformation
\begin{align}
L_{j,\mathbf{m}}(\la \bar{a}^{\nu})=
q^
{-\frac 1 2 \sigma ^3_j\sum_{\mathbf{i}=1}^{\mathbf{m}}\sigma ^3_{\mathbf{i}}}
\widehat{L}_{j,\mathbf{m}}(\la \bar{a}^{\nu})
q^
{\frac 1 2 \sigma ^3_j\sum_{\mathbf{i}=1}^{\mathbf{m-1}}\sigma ^3_{\mathbf{i}}}\,,
\label{gauge}
\end{align}
where
$$
\widehat{L}_{j,\mathbf{m}}(\la \bar{a}^{\nu})=
\begin{pmatrix}
1- \bar{a}^{2\nu}\la ^2
q^{ \sigma ^3_\mathbf{m} }&-(q-q^{-1})\la  \bar{a}^{\nu}
q^
{-\frac 1 2 +\sum_{\mathbf{i}=1}^{\mathbf{m-1}}
\sigma ^3_{\mathbf{i}}}\sigma ^-_\mathbf{m}
\\
-(q-q^{-1})\la  \bar{a}^{\nu}
q^
{-\frac 1 2 -\sum_{\mathbf{i}=1}^{\mathbf{m-1}}
\sigma ^3_{\mathbf{i}}}\sigma ^+_\mathbf{m} &
1- \bar{a}^{2\nu}\la ^2q^{- \sigma ^3_\mathbf{m}}
\end{pmatrix}_j\,.
$$

Now we recall known formulae concerning the continuous limit 
of the XXZ chain \cite{LP}. 
A very accurate account of this matter 
is given in Lukyanov's paper \cite{Luk}. 
Notice, however, that the Hamiltonian in 
\cite{Luk} differs from ours by a similarity 
transformation with the operator 
$U=\prod_{\mathbf{m}:\ \mathrm{odd}}\sigma ^3_\mathbf{m}$. 
Having this in mind we rewrite the 
main order formulae for $\mathbf{n}\to\infty$, 
$y=\mathbf{m} a$ (formulae (2.19) in \cite{Luk}) as
follows. 
\begin{align}
&\sigma ^3_{\mathbf{m}} \ \to\  \frac a{i\pi (1-\nu)}
\partial_y
\(\varphi (-iy)-\bar{\varphi }(-iy)\)\,,
\label{lukbos}\\
&\sigma ^{\pm}_{\mathbf{m}}
\ \to\ (-1)^\mathbf{m}\sqrt{ F /2} \ a^{\frac 1 2(1-\nu)}
:e^{\pm (\varphi(-iy)+\bar{\varphi}(-iy))}:\nn\,.
\end{align}
Here $\varphi (x)$, $\bar{\varphi}(x)$ are two chiral bosons
with the same normalisation as in section \ref{reviewBLZ}. 
The fractional power 
of $a$ in the second formula is needed in order to
compensate the anomalous dimension of 
$:e^{\pm (\varphi(-iy)-\bar{\varphi}(-iy))}:$, and 
$F$ is 
related to the one point function of the latter \cite{Luk}. 
From these formulae we see that
\begin{align}
W^{\pm}_m=q^{ \mp\sum_{\mathbf{i}=1}^{\mathbf{m-1}}
\sigma ^3_{\mathbf{i}}}\sigma ^{\pm}_\mathbf{m}
\ \to\
\sqrt{Z} \ a^{1-\nu}
V_{\pm}(-iy)\,,
\end{align}
where $V_{\pm}(x)$ are the 
chiral vertex operators \eqref{VO}. The power of $a$ changed due to a normal reordering,
while the constant $Z$ is obviously related to
the asymptotical behaviour at $m\to\infty$ of the following 
two-point function for XXZ model:
$$\langle W^+_mW^-_0\rangle=\frac{Z}{m ^{2(1-\nu)}}\,.$$
We do not know a direct way to fix this constant, however,
our construction allows an indirect one. Indeed, 
in order to have complete agreement with CFT on this
microscopic level we need that
\begin{align}
&\widehat{L}_{j,\mathbf{m}}(\la \bar{a}^{\nu})
=1+a \la
\mathcal{K}_{j,\mathbf{H}}(-iy)
+O(a ^{2\nu})\,,
\end{align}
which would imply
\begin{align}
&\widehat{L}_{j,\mathbf{n}}(\la\bar{a}^{\nu})
\cdots 
\widehat{L}_{j,\mathbf{1}}(\la \bar{a}^{\nu})
\ \to\ 
\mathcal{P}\exp
\left(\la\int_0^{2\pi R}
\mathcal{K}_{j,\mathbf{H}}(-iy)dy\right), 
\end{align}
where we have set 
\begin{align*}
&\mathcal{K}_{j,\mathbf{H}}(x)
=\ iq^{-\frac 1 2}
\( \sigma ^+_jV_-(x)+\sigma ^-_jV_+(x)\)
\,.
\end{align*}
So the microscopic picture agrees with the macroscopic one if
$$
Z=\frac 1 {4\sin^2(\pi\nu)C^{2\nu}}\,.
$$
Altogether we obtain from \eqref{TM}
\begin{align}
T_\mathbf{M}(\la \bar{a}^{\nu},\kappa)\ 
\to\  
\Tr_j\bigl[
e^{\pi i \nu\kappa}
\mathcal{P}\exp\bigl(\la\int_0^{2\pi R}
\mathcal{K}_{j,\mathbf{H}}(-iy)dy\bigr)
\bigr], 
\label{limT}
\end{align}
giving rise to the BLZ transfer matrix \eqref{transfer}
with the identification \eqref{nullmode}.  
In particular, the second chirality decouples.

\section{ Conformal limit in the space direction}\label{cftlimitspace}

Let us return to the formula
\begin{align}
&Z^{\kappa,\kappa '}_R
\bigl\{\taub^*(\la^0_1)\cdots \taub^*(\la^0_p)
\betab ^*(\la^+_1)\cdots \betab^*(\la^+_r)
\gammab^*(\la^-_r)\cdots \gammab^*(\la^-_1)
\bigl(\Phi_\al(0)\bigr)
\bigr\}
\label{Zconf}\\
&=\prod\limits _{i=1}^p 
\rhob(\la _i^{0}|\kappa,\kappa')
\det \left(
\omegab(\la^+_i,\la ^-_j|\kappa,\kappa ',\al)
\right)_{i,j=1,\cdots, r}
\,,\nn
\end{align}
We have seen that the functions 
in the right hand side are defined through the eigenvalues of the 
BLZ transfer matrix on 
the primary fields $\phi_{\kappa+1}$, $\phi_{-\kappa'+1}$.
Thus the right hand side of \eqref{Zconf} is defined. We want to interpret the left hand side
of this equation. Our arguments are far from being  mathematically rigourous,
so, we formulate our statement  as a conjecture.

\vskip .3cm
\noindent{\bf Conjecture.}
{\it Asymptotics of \eqref{Zconf} for $\la^{\pm}_i, \la ^{0}_i\to\infty$ describes the expectation values of
descendants
for chiral CFT with
$c=1-6\frac {\nu ^2}{1-\nu}$ of the primary field $\phi _{\al}(0)$ inserted on the cylinder with the asymptotic conditions described
by $\phi _{-\kappa ' +1}$ and $\phi _{\kappa +1}$.}
\vskip .3cm

Recall that we start with $\kappa$, $\kappa '$, $\al$ 
satisfying \eqref{kappaprime}, and then continue analytically. 
The three point function 
of the operators
$\phi_ {-\kappa' +1}(-\infty)$,
$\phi_ {\al}(0)$,
$\phi_ {\kappa +1}(\infty)$ 
does not vanish because \eqref{kappaprime} can be rewritten as
$$
(-\kappa ' +1)+
 \al
+(\kappa+1)=2-2\textstyle{\frac{1-\nu
}{\nu}} s\,,$$
which coincides in our normalisation with 
the Dotsenko-Fateev condition \cite{DF} for one type of screening operators condition. 
We do not  know if  the second set of screening operators can be defined starting from the lattice model.

In the present section we shall first present qualitative arguments in favour of this conjecture, and then explain how it can be verified quantitatively. The actual verification will be done in sections 
\ref{asymp-omega}, \ref{final}.

Consider the primary field $q^{2\al S(0)}$ on the lattice. Making the scaling limit in horizontal
direction on the cylinder in the same way as it was done in the vertical one we conclude
that this operator turns into 
$\Phi_\al(0)=\phi _{\al}(0)\otimes \bar{\phi} _{\al} (0)$. 
Typical operators in the space 
$\mathcal{W}_{\al-s,s}$ are of the form
$q^{2(\al-s)S(0)}\sigma ^+_{k_1}\cdots 
\sigma ^+_{k_s}\mathcal{O}$ ($s\ge0$) or 
$q^{2(\al-s)S(0)}\sigma ^-_{k_1}\cdots 
\sigma^-_{k_{-s}}\mathcal{O}$ ($s<0$) 
where $\mathcal{O}$ is spinless.  
Then the same bosonisation formulae as \eqref{lukbos}
$$q^{2\al S(0)}\ \sim \  e^{\frac\nu{1-\nu}\al(\varphi(0)-\bar{\varphi}(0))},
\quad q^{-2S(k-1)}\sigma ^+_k\ \sim \ e^{2\varphi (x)}$$
imply that 
$$\mathrm{Scaling}\ \mathrm{ limit}
\(\mathcal{W}_{\al-s,s}\)\subset
\mathcal{V}_{\al+2\frac{1-\nu }{\nu}s}
\otimes \overline{\mathcal{V}}_{-\al}\,.$$
So, the operators 
$\taub^*(\la)$, $\betab^*(\la)$, $\gammab^*(\la)$
do not change the Verma module for the second chiarlity. 
This is one reason to assume that they do not act on it at all. 
Let us give the first evidence for this claim.

Consider the operator $\taub^*(\la)$. 
It originates from its counterpart on the lattice, $\tb ^*(\z)$. 
We know that close to $\z ^2=1$ the operator $\tb ^*(\z)$ 
describes the 
adjoint action of XXZ local integrals of motion. So, we expect the same in the CFT.
Using the BLZ formulae \eqref{asymptT} we see that
\begin{align}
\log\rhob (\la |\kappa,\kappa')
\simeq\sum\limits _{n=1}^{\infty}\la ^{-\frac {2n-1}\nu}
C_n
\(I_{2n-1}(\kappa)-I_{2n-1}(\kappa' )\)\,,
\label{BLZ-logT}
\end{align}
which implies together with \eqref{Int-m} that 
$$\taub ^*(\la)\simeq \exp\Bigl( \sum\limits _{n=1}^{\infty}
\la ^{-\frac {2n-1}\nu}
C_n
\mathbf{i}_{2n-1}\Bigr)\,.$$
So, as it has been expected,  
the local operators are extracted from the action
of $\taub ^*(\la)$ in the asymptotics at $\la \to \infty$ as coefficients of
fractional degrees  $\la ^{\frac 1 \nu}$ which has the dimension of inverse length.
Certainly, this exercise is quite tautological, and we  would not write this paper if this were
the only thing we can do. But it demonstrates the chiral nature of our operators.

In the appendix we prove the general statement
that $\omegab(\la,\mu |\kappa,\kappa ',\al)$ has the following
asymptotics
for $\la^2,\mu^2\to+\infty$
\begin{align}
&\omegab(\la,\mu |\kappa,\kappa ',\al)
\simeq 
\sqrt{\rhob(\la|\kappa,\kappa')}
\sqrt{\rhob(\mu|\kappa,\kappa')}
\sum\limits_{i,j=1}^{\infty}
\la ^{-\frac{2i-1}\nu}
\mu ^{-\frac{2j-1}\nu} 
\omega_{i, j}(\kappa,\kappa ',\al)\,.
\label{asymom}
\end{align}

The proof will be given for the primary fields $\phi _{-\kappa'-1}$, $\phi _{\kappa +1}$
as asymptotical conditions, but it can be generalised to arbitrary descendants. 
So, in the weak sense the following operators are defined:
\begin{align}
&\log\(\taub^*(\la)\)\simeq
\sum\limits _{j=1}^{\infty}
\taub^*_{2j-1}\la^{-\frac{2j-1}\nu}
\,,
\label{asymt}\\
&\frac 1{\sqrt{\taub^*(\la)}}\betab^*(\la)\simeq\sum\limits _{j=1}^{\infty}\betab^*_{2j-1}\la^{-\frac{2j-1}\nu}
\,,\label{asymb}\\
&\frac 1{\sqrt{\taub^*(\la)}}\gammab^*(\la)
\simeq
\sum\limits _{j=1}^{\infty}\gammab^*_{2j-1}\la^{-\frac{2j-1}\nu}
\,.\label{asymc}
\end{align}
which act between different  Verma modules:
\begin{align}
&\taub^*_{2j-1}\ \ :\  
\mathcal{V}_{\al+2\frac{1-\nu }{\nu}s}\ \ \ \ 
\to\  
\mathcal{V}_{\al+2\frac{1-\nu}{\nu}s}
\,,
\nn\\
&\betab ^*_{2j-1}\ \ :\  
\mathcal{V}_{\al+2\frac{1-\nu}{\nu}(s-1)} \ 
\to\  
\mathcal{V}_{\al+2\frac{1-\nu}{\nu}s}\,,
\nn\\ 
&\gammab^*_{2j-1}\ \ :\  
\mathcal{V}_{\al+2\frac{1-\nu}{\nu}(s+1)} \ 
\to\  
\mathcal{V}_{\al+2\frac{1-\nu}{\nu}s}\,.
\nn
\end{align}

Consider the Verma module 
$\mathcal{V}_{\al}$. 
We obtain different elements of this module by operators $\taub^*_{2j-1}$ acting on the primary field 
$\phi_{\al}$ , and by the same number of operators $\betab^*_{2j-1}$ and $\gammab^*_{2j-1}$ acting further. 
Due to the completeness in the lattice case
\cite{complet}, in this way we obtain
linearly independent vectors from $\mathcal{V}_{\al}$. 
Counting the characters we see that
the entire Verma module is created in this way. 
Indeed, from a combinatorial point of view,  
$\taub^*$ is one odd boson, and $\betab^*$, $\gammab^*$
are Gross-Neveu fermions 
which in uncharged sector produce one even boson. 

No we shall proceed to computation
of the coefficients $\omega_{i, j}(\kappa,\kappa ',\al)$ and comparing them
with the three-point functions of CFT. Since the operators $\taub^*(\la)$ is
already settled we shall consider the case $\kappa =\kappa '$ which means to
ignore the image of the actions by the local integrals of motion $\mathbf{i}_{2n-1}$
in the Verma module $\mathcal{V}_\al $.  Acting on the primary field $\phi _{\al}(0)$,
the even generators of the Virasoro algebra, $\mathbf{l}_{-2k}$, create
the quotient space of the Verma module by these descendants.

\section{Asymptotics of 
$\log \mathfrak{a}^\mathrm{sc}$}
\label{asymp-loga}

In this section we study the asymptotic behaviour of the
function 
\begin{align*}
\frak{a}^\mathrm{sc}(\la,\kappa)=\frac{\Qb(\la q,\kappa )}
{\Qb(\la q ^{-1},\kappa)}
\end{align*}
as $\la^2\to\infty$.   
Following closely the analysis developed in \cite{BLZII},    
we give a recursive algorithm for 
determining the coefficients of the asymptotic expansion. 

It is known (see \cite{BLZII}, (3.17) and (3.23)) 
that for large $\kappa$ 
the smallest Bethe root behaves as 
$\lambda_1^2\sim c(\nu)\kappa^{2\nu}$, where
\begin{align}
&c(\nu)=
\Gamma(\nu)^{-2}e^\delta \left(\frac{\nu}{2R}\right)^{2\nu},
\label{delta}\\
&\delta=-\nu\log\nu-(1-\nu)\log(1-\nu)\,.
\nn
\end{align}
The main technical idea in \cite{BLZII} is to consider the limit 
\begin{align*}
\la^2,\ \kappa\to\infty\,,
\qquad \mathrm{keeping}\ 
\ 
t=c(\nu)^{-1}\frac{\lambda^2}{\kappa^{2\nu}}
\ \ \mathrm{ fixed}. 
\end{align*}

Henceforth we change the variable from $\lambda$ to 
$t$ and write 
\begin{align*}
F(t,\kappa)=\log\mathfrak{a}^\mathrm{sc}(\la,\kappa)\,.
\end{align*}
This function is to be determined from the 
non-linear integral equation.
In order to write the equation, it is convenient 
to redefine some functions in terms of the variables $t,u$.
We use
\begin{align*}
&K(t)=\frac1{2\pi i}\cdot\frac12\left(\frac{tq^2+1}{tq^2-1}
-\frac{tq^{-2}+1}{tq^{-2}-1}\right).
\end{align*}
We also use $R(t,u)$ to represent the following resolvent 
kernel.
\begin{align*}
&R(t,u)-\int\limits_1^\infty\frac{dv}vR(t,v)K(v/u)=K(t/u),
\quad(t,u>1).
\end{align*}
An explicit formula for $R(t,u)$ will be  given below.

The non-linear integral equation for $F(t,\kappa)$ $(t>1)$ reads\\
\begin{align}
&F(t,\kappa)-
\int\limits_{1}^\infty K(t/u)F(u,\kappa)\frac{du}{u}
=-2\pi i \nu\kappa 
\label{DDVsc}\\
&-
\Bigl(\int\limits_{1}^{e^{i\epsilon}\cdot\infty}
K(t/u)\log(1+e^{F(u,\kappa)})\frac{du}{u}
-\int\limits_{1}^{e^{-i\epsilon}\cdot\infty}
K(t/u)\log(1+e^{-F(u,\kappa)})\frac{du}{u}
\Bigr)\,,\nn
\end{align}
where $\epsilon$ is a small positive number.
From Appendix A, we see that 
\begin{align*}
&F(t,\kappa)=
-F_+(\arg t,\kappa)|t|^{\frac1{2\nu}}+O(|t|^{-\frac1{2\nu}}),\quad0<\arg t<\pi,\\
&F(t,\kappa)=\phantom{-}
F_-(\arg t,\kappa)|t|^{\frac1{2\nu}}+O(|t|^{-\frac1{2\nu}}),\quad-\pi<\arg t<0,
\end{align*}
where $F_\pm$ is positive. We seek for the solution of 
\eqref{DDVsc} in an asymptotic 
series in $\kappa^{-1}$, 
\begin{align*}
F(t,\kappa)\simeq \sum_{n=0}^{\infty}\kappa^{-2n+1}F_n(t)\,.
\end{align*}

Consider first the leading coefficient $F_0(t)$. 
For $t>1$, from (9.2) follows
\begin{align}
((I-K)F_0)(t)=-2\pi i\nu,
\label{WiHp0}
\end{align}
where $K$ denotes the integral 
operator on the interval $[1,\infty)$ 
\begin{align*}
Kf(t)=\int\limits_{1}^\infty K(t/u)f(u)\frac{du}{u}\,.
\end{align*}
Equation \eqref{WiHp0} can be 
solved by the standard Wiener-Hopf technique. 

Quite generally, for a function $f(t)$ let 
\begin{align*}
\hat{f}(k)=\int\limits_0^\infty f(t)t^{-ik}\frac{dt}{t},
\qquad 
f(t)=\int\limits_{-\infty}^\infty \hat{f}(k)t^{ik}
\frac{dk}{2\pi}\,
\end{align*}
denote the Mellin transform and its inverse transform. 
For the solution of \eqref{WiHp0} we shall need
\begin{align*}
&\hat{K}(k)=\frac{\sinh(2\nu-1)\pi k}{\sinh\pi k}\,,
\end{align*}
along with the Riemann-Hilbert factorisation
\begin{align*}
&1-\hat{K}(k)=S(k)^{-1}S(-k)^{-1}\,,
\\
&S(k)=\frac{\Gamma(1+(1-\nu)ik)\Gamma(1/2+i\nu k)}
{\Gamma(1+ik)\sqrt{2\pi(1-\nu)}}e^{i\delta k}\,,
\end{align*}
where $\delta$ is defined in \eqref{delta}.

The function $S(k)$ is holomorphic on the lower half plane
$\Im k<1/2\nu$,  
and for $k\to\infty$ behaves as $S(k)=1+O(k^{-1})$.  
If we demand that 
\begin{align*}
F_0(t)=const.\ t^{\frac{1}{2\nu}}
+O\left(t^{-\frac{1}{2\nu}}\right)
\qquad (t\to\infty)\,,
\end{align*}
then \eqref{WiHp0} admits a unique solution given by 
\begin{align}
&F_0(t)=
\int\limits_{\R-\frac{i}{2\nu}-i0}
dl\,t^{il}S(l)\frac{-if}{l(l+\frac{i}{2\nu})}\,,
\qquad (t>1)\,
\label{F0}
\end{align}
where
\begin{align*}
&f=\frac{1}{2\sqrt{2(1-\nu)}}\,.
\end{align*}
The right hand side of (9.4) gives a continuous function $\tilde F_0(t)$ on the half line $(0,\infty)$
such that $\tilde F_0(t)=0$ for $0<t\leq1$. However, the function $F_0(t)$ for $t>1$ can be analytically continued
to the sector $|\arg t|<2(1-\nu)\pi$, rewriting the equation (9.3): 
\begin{align}
&F_0(t)=-2\pi i\nu+(KF_0)(t)\,,
\label{F00}\\
&(KF_0)(t)=\int\limits_{\R-\frac{i}{2\nu}-i0}
dl\,t^{il}S(l)\hat{K}(l)
\frac{-if}{l(l+\frac{i}{2\nu})}\,.
\label{F000}
\end{align}
It is also possible to check directly the consistency of the formulas \eqref{F0} and \eqref{F00}.
The difference of two integrals \eqref{F0} and \eqref{F000} has the only pole in the upper half
plane ${\rm Im}\,l+\frac1{2\nu}>0$ at $l=0$, where we pick up the residue $-2\pi i\nu$.

Now we turn to the higher order terms. 
Similarly as above, the Wiener-Hopf method  
allows us to find $R(t,u)$ for $t>1$,
\begin{align*}
&R(t,u)=
\int\limits_{-\infty}^\infty\!\!
\int\limits_{-\infty}^\infty\frac{dl}{2\pi}\frac{dm}{2\pi}
t^{il}u^{im}S(l)S(m)\hat{K}(m)
\frac{-i}{l+m-i0}\,.
\end{align*}
The analytic continuation is given by
\begin{align*}
&R(t,u)=
K(t/u)+
\int\limits_{-\infty}^\infty\!\!
\int\limits_{-\infty}^\infty\frac{dl}{2\pi}\frac{dm}{2\pi}
t^{il}u^{im}S(l)S(m)\hat{K}(l)\hat{K}(m)
\frac{-i}{l+m-i0}
\,.
\end{align*}
From this follows
\begin{align*}
&R(t,u)=
\int\limits_{-\infty}^\infty\frac{dl}{2\pi}
t^{il}S(l)\hat{K}(l)\hat R(l,u)\,,
\end{align*}
where
\begin{align}
\hat R(l,u)&=
\int\limits_{-\infty}^\infty\frac{dm}{2\pi}
u^{im}S(m)\frac{-i}{l+m-i0}\label{Rtu}\\
&=u^{-il}S(l)^{-1}+
\int\frac{dm}{2\pi i}\frac{1}{l+m-i0}
u^{im}S(m)\hat{K}(m).\nonumber
\end{align}
The last line shows that $\hat R(l,e^x)$ is analytic near $x=0$.
Equation \eqref{DDVsc} can be converted into 
\begin{align}
&F(t,\kappa)
=\kappa F_0(t)
\label{DDV2}\\
&-
\Bigl(\int\limits_{1}^{e^{i\epsilon}\cdot\infty}
R(t,u)\log(1+e^{F(u,\kappa)})\frac{du}{u}
-\int\limits_{1}^{e^{-i\epsilon}\cdot\infty}
R(t,u)\log(1+e^{-F(u,\kappa)})\frac{du}{u}
\Bigr)\,.\nn
\end{align}

Motivated by the formula \eqref{F00}, let us set 
\begin{align}
&F(t,\kappa)=\kappa F_0(t)+
\int\limits_{-\infty}^\infty
dl\,t^{il}S(l)\hat{K}(l)(\Psi(l,\kappa)-
\kappa\Psi_0(l))\,,\label{F}
\end{align}
where $\Psi(l,\kappa)$ has an asymptotic expansion, 
\begin{align}
&\Psi(l,\kappa)\simeq
\sum_{n=0}^\infty\kappa^{-2n+1}\Psi_n(l)\,,
\quad 
\Psi_0(l)=\frac{-if}{l(l+\frac{i}{2\nu})}\,.
\label{asyPsi}
\end{align}
We show below that each coefficient $\Psi_n(l)$ ($n\ge1$) 
can be determined as a polynomial in $l$  
by a purely algebraic procedure.  

With a change of integration variable, 
\eqref{DDV2} is brought further into the form 
\begin{align}
\Psi(l,\kappa)-\kappa\Psi_0(l)
=-\frac{i}{f\kappa}
\Bigl\{
\int\limits_{0}^{-i\infty+\epsilon}\frac{dx}{2\pi}
&\hat{R}(l,e^{ix/f\kappa})\log(1+e^{F(e^{ix/f\kappa},\kappa)})
\label{DDV3}\\
+\int\limits_{0}^{i\infty+\epsilon}\frac{dx}{2\pi}
&\hat{R}(l,e^{-ix/f\kappa})\log(1+e^{-F(e^{-ix/f\kappa},\kappa)})
\Bigr\}\,.
\nn
\end{align}
Since $\log(1+e^{\pm F(u,\kappa)})$ decays exponentially
for $\pm\Im u>0$, the asymptotics of the right hand side 
of \eqref{DDV3} is completely determined from the 
behaviour of the integrand at $x=0$. 

In order to develop a systematic expansion,  
let us first make a general remark. 
Consider a Fourier integral 
\begin{align*}
G(x)=\int_{-\infty}^{\infty}e^{ikx}g(k) dk\,.
\end{align*}
We assume that $g(k)$ is the boundary value of a 
holomorphic function on the lower half plane 
$\mathop{\rm \Im} k<0$, satisfying the asymptotic expansion 
\begin{align*}
g(k)\simeq \sum_{n=-n_0}^\infty g_n (ik)^{-n}\quad 
(k\to \infty,\ \mathop{\rm Im}k<0). 
\end{align*}
Then integration by parts shows that for any $N>0$ we have 
\begin{align*}
G(x)=\sum_{n=-n_0}^0 g_n 2\pi \delta^{(n)}(x)+
2\pi \sum_{n=1}^{N}\frac{g_n }{(n-1)!} x_+^{n-1}+R_N,
\end{align*}
where $R_N=O(x^N)$ as $x\to 0$. 
Suppose further that $G(x)$ can be prolonged 
analytically around $x=0$. 
In this situation its Taylor expansion can be computed 
from the right hand side,  
discarding the delta function terms. 
The result is summarised in a compact form 
\begin{align*}
\int_{-\infty}^{\infty}e^{ikx}g(k) dk
=2\pi i\,{\rm res}_{k}[e^{ikx}g(k)]\,,
\end{align*}
where $\res_k\bigl[\cdots\bigr]$ 
signifies the coefficient of 
$k^{-1}$ in the expansion at $k=\infty$.  

The above consideration applies to \eqref{Rtu}, and 
we obtain the Taylor expansion at $x=0$, 
\begin{align}
\hat{R}(l,e^{ix/f\kappa})=
\res_{h}\Bigl[\frac{e^{-h x/f\kappa}}{l+h}S(h)\Bigr]\,.
\label{Rhat}
\end{align}

For the factor $\log(1+e^{F(u,\kappa)})$, we proceed as follows.
Set 
\begin{align*}
&F(e^{ix/f\kappa},\kappa)
=-2\pi\left(x-\bar{F}(x,\kappa)\right)\,.
\end{align*}
Similarly as above, 
the Taylor expansion of $\bar{F}(x,\kappa)$ at $x=0$ is calculated as 
\begin{align}
&\bar{F}(x,\kappa)= x+\res_{h}
\Bigl[e^{-h x/f\kappa}S(h)i\Psi(h,\kappa)\Bigr]\,.
\label{TayF}
\end{align}
Actually the term $x$ is cancelled by 
a term coming from $\kappa\Psi_0(h)$, 
so that $\bar{F}(x,\kappa)=O(\kappa^{-1})$. 
We rewrite the corresponding piece of the integrand as 
\begin{align}
&\log(1+e^{\pm F(e^{\pm ix/f\kappa},\kappa)})
=\sum_{n=0}^\infty 
\frac{\bar{F}(\pm x,\kappa)^n}{n!}
\Bigl(\mp \frac{\partial}{\partial x}\Bigr)^n
\log(1+e^{-2\pi x})\,. 
\label{Fbar}
\end{align}
Substituting \eqref{Fbar}, \eqref{Rhat} into \eqref{DDV3}, 
we arrive at 
\begin{align}
&i\Psi(l,\kappa)-i\kappa \Psi_0(l)
\label{recF}\\
&\simeq 
\frac{2}{f\kappa}\sum_{n=0}^\infty \frac{1}{n!}
\int\limits_{0}^\infty\frac{dx}{2\pi}
\left\{\res_{h}\Bigl[\frac{e^{-h x/f\kappa}}{l+h}S(h)\Bigr]
\bar{F}(x,\kappa)^n
\Bigl(-\frac{\partial}{\partial x}\Bigr)^n\right\}_{\rm even}
\log(1+e^{-2\pi x})\,.
\nn
\end{align}
Here $\{\cdots\}_{\mathrm{even}}$ 
(resp. $\{\cdots\}_{\mathrm{odd}}$) 
means
the even (resp. odd) part in $x$. 	
To evaluate the integral in \eqref{recF}
we need only to develop the integrand into a Taylor series and  
apply the formula 
\begin{align*}
&\int\limits_0^\infty\frac{dx}{2\pi}x^m
\Bigl(-\frac{\partial}{\partial x}\Bigr)^n
\log(1+e^{-2\pi x})
=m!(1-2^{-m-1+n})
\frac{\zeta(m-n+2)}{(2\pi)^{m-n+2}}\,.
\end{align*}

In summary, the asymptotic expansion \eqref{asyPsi}
can be calculated order by order in $\kappa^{-2}$, 
from the set of equations \eqref{recF} and \eqref{TayF}. 

The first few terms of the expansion read
\begin{align*}
&i\Psi(l,\kappa)
=\frac{1}{l(l+\frac{i}{2\nu})}f\kappa
+\frac{1}{24}\frac{1}{f\kappa}
\\
&\quad 
+\frac{7}{2^6\cdot 90}
\left(l-\frac{i}{2\nu}\right)
\left(l-i\frac{2\nu^2-6\nu+6}{7\nu(1-\nu)}\right)
\frac{1}{(f\kappa)^3}
+\cdots
\end{align*}
In general, the coefficients have the structure
\begin{align}
&\Psi_n(l)=\prod_{j=1}^{n-1}\left(l-\frac{i(2j-1)}{2\nu}\right)
\times (\text{ Polynomial in $l$ of degree $ n-1$})\,.
\label{FactorPsi}
\end{align}

From the knowledge of 
$\log\mathfrak{a}^\mathrm{sc}(\la,\kappa)$,  
it is straightforward to extract the asymptotic 
expansion of $\log \Tb(\la,\kappa)$ 
\cite{BLZII}:

\begin{align*}
\log \Tb(\la,\kappa)
&\simeq 
\pi i \nu \kappa+
\frac{\sqrt{1-\nu}}{\sqrt{2\pi}}
\int_{\R-\frac i{2\nu}-i0}dl\ 
\frac{\Gamma(1-il)\Gamma(\frac{1}{2}+i\nu l)}
{\Gamma(1-i(1-\nu)l)}\Psi(l,\kappa)
\Bigl(\frac{e^{\delta-\pi i\nu}\la^2}{\kappa^{2\nu}c(\nu)}\Bigr)^{il}
\\&
\simeq 
\sum_{n=0}^\infty
C_n I_{2n-1}(\kappa) 
\la ^{-\frac {2n-1}{\nu}}
\end{align*}
where
\begin{align}
&C_n=-\frac{\sqrt{\pi}}{\nu}\frac{1}{n!}\frac{\Gamma(\frac{2n-1}{2\nu})}{\Gamma(1+\frac{1-\nu}{2\nu}(2n-1))}
(1-\nu)^n 
\ \Gamma(\nu)^{-\frac {2n-1}{\nu}}\,,
\label{Ikappa}
\\
&I_{2n-1}(\kappa)=-i\Psi\left(\frac{i(2n-1)}{2\nu},\kappa\right)
\times n(2n-1)(2\nu^2)^{n-1}(f\kappa)^{2n-1}
R^{-2n+1}\,,
\nn
\end{align}
and we set by definition $I_{-1}=R$. 
Notice that $C_0>0$ while $C_n<0$ for $n\ge 1$.

The factors in \eqref{FactorPsi} ensure
that at these special values of $l$ 
the asymptotic series \eqref{asyPsi} truncates. 
This has to be the case, because 
according to \cite{BLZII} 
$I_{2n-1}(\kappa)$ ($n\ge1$) are the vacuum eigenvalues 
of the integrals of motion which are  
polynomials in $c$ and $\Delta_{\kappa+1}$.

For instance 
\begin{align}
&I_1(\kappa)=\frac{1}{R}\Bigl(
\Delta_{\kappa+1}-\frac{c}{24}\Bigr),
\label{I1}\\
&I_3(\kappa)=\frac{1}{R}I_1(\kappa)^2-
\frac{1}{6R^2}I_1(\kappa)
+\frac{c}{1440R^3}\,,
\label{I3}
\\
&I_5(\kappa)=\frac{1}{R}
I_3(\kappa)I_1(\kappa)-
\frac{1}{3R^2}I_3(\kappa)
+\frac{c+5}{360R^4}
I_1(\kappa)
-\frac{c(5c+28)}{181440R^5}\,.
\label{I5}
\end{align}
We have verified upto $n=4$ that \eqref{Ikappa}
matches perfectly 
the formulas for $I_{2n-1}$ given in \cite{BLZI}.

\section{Asymptotics of 
$\omega$ for $\kappa=\kappa'$}
\label{asymp-omega}

In this section, we restrict our consideration to the case 
$\kappa=\kappa'$, so that 
$\rhob(\la|\kappa,\kappa')=1$. 
Our goal is to give an algorithm for deriving 
the asymptotic expansion of the function 
$\omegab(\la,\mu|\kappa,\kappa, \al)$. 

We start from the representation 
\begin{align}
&\omegab(\la,\mu|\kappa,\kappa,\alpha)
=\Bigl(
f_\mathrm{left}\star  f_\mathrm{right}+
f_\mathrm{left}\star 
\Rdr
\star f_\mathrm{right}\Bigr)(\la,\mu)+\omega_0(\la,\mu|\al)\,,
\label{FRF}
\end{align}
where
\begin{align*}
&f_\mathrm{left}(\la,\mu,\alpha)=
\frac{1}{2\pi i}
\delta_\lambda^-\psi(\lambda/\mu,\alpha),
\quad 
f_\mathrm{right}(\la,\mu,\alpha)=\delta_\mu^-\psi(\lambda/\mu,\alpha)\,,\nn\\
&\omega _0(\la,\mu|\al)
=\delta_\lambda^-\delta_\mu^-\Delta ^{-1}_\la 
\psi(\lambda/\mu,\alpha)\,,
\end{align*}
and $\Rdr$ denotes the resolvent for the integral equation 
\begin{align}
\Rdr
-   \Rdr    \star K_\alpha  
=K_\alpha\,. 
\label{dress}
\end{align} 
Here we have set 
\begin{align*}
&f \star g=
\left(\int_{\sigma^2}^{e^{i\epsilon}\infty}
-\int_{\sigma^2}^{e^{-i\epsilon}\infty}\right)
f(\la)g(\la)dm(\la)\,,
\\
&dm(\la)=
\frac{d\la^2}{\la^2
(1+\mathfrak{a}^\mathrm{sc}(\la,\kappa))}\,,
\end{align*}
and $\sigma^2$ is a point lying 
between the smallest Bethe 
root and the largest zero of $\Tb(\la,\kappa)$. 
Strictly speaking, $\star$ and $dm(\la)$
have slightly different meaning than those 
used in section \ref{rhomega}. Since we use them 
here only locally, there should not be a fear of confusion. 

From now untill \eqref{omth} below, 
we shall work with the variables 
\begin{align*}
&t=c(\nu)^{-1}\lambda^2/\kappa^{2\nu},  
\quad
u=c(\nu)^{-1}\mu^2/\kappa^{2\nu}\,, 
\end{align*}
and write 
\begin{align*}
K(t,\al)=
\frac1{2\pi i}\cdot\frac12
\left((tq^2)^{\al/2}\frac{tq^2+1}{tq^2-1}
-(tq^{-2})^{\al/2}\frac{tq^{-2}+1}{tq^{-2}-1}\right).
\end{align*} 

Again we start with the leading order 
approximation as $\kappa\to\infty$,   
where \eqref{dress} becomes 
\begin{align*}
R(t,u,\alpha)-\int_1^\infty \frac{dv}{v}
R(v,u,\alpha)K(t/v,\alpha)=K(t/u,\alpha)\,.
\end{align*}
This can be solved in the same manner as before, using 
\begin{align*}
&\hat{K}(k,\alpha)=
\frac{\sinh\pi\bigl((2\nu-1)k-\frac{i\alpha}{2}\bigr)}
{\sinh\pi\bigl(k+\frac{i\alpha}{2}\bigr)}\,. 
\end{align*}
The only point worth noting is that in writing 
the Riemann-Hilbert 
factorisation
\begin{align*}
&1-\hat{K}(k,\alpha)=S(k,\alpha)^{-1}S(-k,2-\alpha)^{-1},
\\
&S(k,\alpha)=
\frac{\Gamma\bigl(1+(1-\nu)ik-\frac{\alpha}{2}\bigr)
\Gamma\bigl(\frac{1}{2}+i\nu k\bigr)}
{\Gamma\bigl(1+ik-\frac{\alpha}{2}\bigr)
\sqrt{2\pi}(1-\nu)^{(1-\alpha)/2}}e^{i\delta k}\,,
\end{align*}
we are naturally led to assume that 
\begin{align*}
0<\alpha<2\,. 
\end{align*}
So the na{\"\i}ve symmetry 
$(-k,-\alpha)\to(k,\alpha)$ of $\hat{K}(k,\alpha)$ 
is replaced by the symmetry $(k,\alpha)\to (-k,2-\alpha)$. 
With our normalisation of $\al$, the reflection 
$\al \to 2-\al$ is the usual one for CFT with $c<1$.
For the resolvent kernel we obtain the representation
\begin{align*}
&R(t,u,\alpha)=K(t/u,\alpha)
\\
&+\int_{-\infty}^\infty\!\!
\int_{-\infty}^\infty\frac{dl}{2\pi}\frac{dm}{2\pi}
t^{il}u^{im}S(l,\alpha)
S(m,2-\alpha)\hat{K}(l,\alpha)\hat{K}(m,2-\alpha)
\frac{-i}{l+m-i0}\,.
\end{align*}

The `dressed' resolvent kernel $\Rdr(t,u)$ 
satisfies
\begin{align*}
&\Rdr(t,u)
-R(t,u,\alpha)
\\
&=
-\int_1^{e^{i\epsilon}\infty}\frac{1}
{1+e^{-F(v,\kappa)}
}
R(t,v,\alpha)
\Rdr(v,u)
\frac{dv}{v}
-\int_1^{e^{-i\epsilon}\infty}
\frac{1}{1+e^{F(v,\kappa)}}
R(t,v,\alpha)
\Rdr(v,u)
\frac{dv}{v}\,.
\end{align*}
Setting 
\begin{align}
&\Rdr(t,u)=
K(t/u,\alpha)
\\
&\quad
+\int_{-\infty}^\infty\!\!
\int_{-\infty}^\infty\frac{dl}{2\pi}\frac{dm}{2\pi}
t^{il}u^{im}S(l,\alpha)S(m,2-\alpha)\hat{K}(l,\alpha)
\hat{K}(m,2-\alpha)\Theta(l,m|\kappa,\alpha)\,,
\nn\\
&\Theta(l,m|\kappa,\alpha)
\simeq
\sum_{n=0}^\infty \Theta_n(l,m|\alpha)\kappa^{-2n}\,,
\qquad \Theta_0(l,m)=\frac{-i}{l+m}\,,
\label{asyTh}
\end{align}
and repeating the analysis of the previous section, 
we arrive at the linear recursion relation for the 
$\Theta_n(l,m|\alpha)$: 
\begin{align*}
&\Theta(l,m|\kappa,\alpha)-\Theta_0(l,m)
\simeq
\frac{2}{f\kappa}\sum_{n=0}^\infty \frac{1}{n!}
\int\limits_{0}^\infty dx
\Bigl\{
\res_{l'}\Bigl[\frac{e^{-l' x/f\kappa}}{l+l'}S(l',2-\alpha)\Bigr]
\\
&\times
\res_{m'}\Bigl[e^{-m' x/f\kappa}S(m',\alpha)
\Theta(m',m|\kappa,\alpha)
\Bigr]
\bar{F}(x,\kappa)^n
\Bigl(-\frac{\partial}{\partial x}\Bigr)^n\Bigr\}_{\rm odd}
\times \frac{1}{1+e^{2\pi x}}\,.
\end{align*}
The coefficients of the series \eqref{asyTh}
can be calculated by Taylor expanding the integrand and applying
\begin{align*}
&\int\limits_0^\infty
\,dx x^m
\Bigl(-\frac{\partial}{\partial x}\Bigr)^n
\frac{1}{1+e^{2\pi x}}
=m!(1-2^{-m+n})
\frac{\zeta(m-n+1)}{(2\pi)^{m-n+1}}\,.
\end{align*}
The first non-trivial term reads
\begin{align*}
i\Theta(l,m|\kappa,\alpha)
\simeq
\frac{1}{l+m}+\frac{i}{24\nu}\frac{1}{(f\kappa)^2}
\left(-i\nu(l+m)-\frac{1}{2}+\Delta_\alpha\right)+
O\left(\frac{1}{\kappa^4}\right)\,.
\end{align*}

Returning to the original
variables $\la$ and $\mu$, 
formula for $\omegab(\la,\mu|\kappa,\kappa,\alpha)$ 
can be obtained from \eqref{FRF}.  
We have 
\begin{align}
&\omegab(\la,\mu|\kappa,\kappa,\alpha)
\simeq
\frac{1}{2\pi i}
\int\!\!\int dl dm
\tilde{S}(l,\alpha)
\tilde{S}(m,2-\alpha)
\Theta(l+i0,m|\kappa,\alpha)\,
\nn\\
&\qquad\qquad \qquad\qquad\qquad\quad\times
\Bigl(\frac{e^{\delta+\pi i\nu}\la^2}
{\kappa^{2\nu}c(\nu)}\Bigr)^{il}
\Bigl(\frac{e^{\delta+\pi i\nu}\mu^2}
{\kappa^{2\nu}c(\nu)}\Bigr)^{im}\,,
\label{omth}\\
&\tilde{S}(k,\alpha)=
\frac{\Gamma\bigl(-ik+\frac{\alpha}{2}\bigr)
\Gamma\bigl(\frac{1}{2}+i\nu k\bigr)}
{\Gamma\bigl(-i(1-\nu)k+\frac{\alpha}{2}\bigr)
\sqrt{2\pi}(1-\nu)^{(1-\alpha)/2}}
\,,
\nn
\end{align}
where $l+i0$ is important only in $\Theta _0(l,m)$. 
Notice that originally in $R(t,u,\al)$ we
had rather $l-i0$. 
The change appeared due to addition of 
$\omega_0(\la,\mu)$
which explains the importance of this term. 
Picking the residues at the poles in the upper half plane,  
its asymptotics as $t,u\to\infty$ can be calculated: 
\begin{align}
\omegab(\la,\mu|\kappa,\kappa,\alpha)
&\simeq 
\sum_{r,s=1}^\infty
\frac{1}{r+s-1}
D_{2r-1}(\alpha)D_{2s-1}(2-\alpha)
\label{asym-omega}
\\
&
\quad\times
\la^{-\frac{2r-1}\nu} 
\mu ^{-\frac{2s-1}\nu} 
\Omega_{2r-1,2s-1}(\kappa,\alpha)\,,
\nn
\end{align}
where
\begin{align}
&D_{2n-1}(\alpha)=
\frac 1{\sqrt{i\nu}}
\ \Gamma (\nu)^{-\frac{2n-1}\nu}(1-\nu)^{\frac{2n-1}{2}}
\cdot \frac{1}{(n-1)!}
\frac{\Gamma\left(\frac{\alpha}{2}+\frac 1{2\nu}(2n-1)\right)}
{\Gamma\left(\frac{\alpha}{2}+\frac{(1-\nu)
}{2\nu}(2n-1)
\right)}
\,,
\label{Phi}
\end{align}
and 
\begin{align*}
\Omega_{2r-1,2s-1}(\kappa,\alpha)
&=-\Theta\left(\frac{i(2r-1)}{2\nu} ,\frac{i(2s-1)}{2\nu}
\Bigl|
\kappa,\alpha\right)
\times \frac{r+s-1}{\nu}
\left(\frac{\sqrt{2} \ f\kappa \nu}{R}\right)^{2r+2s-2}\,.
\end{align*}
The counterpart of the factorisation \eqref{FactorPsi} for 
$\Psi_n(l)$ 
is the vanishing property 
\begin{align*}
\Theta_n\left(\frac{i(2r-1)}{2\nu} ,\frac{i(2s-1)}{2\nu}
\Bigl|\alpha\right)=0
\qquad (n\ge r+s).
\end{align*}
This ensures that the 
coefficients $\Omega_{2r-1,2s-1}(\kappa,\alpha)$ are 
polynomials in $\Delta_{\kappa+1}$, $\alpha$ and $c$. 
For instance, 
\begin{align*}
&\Omega_{1,1}(\kappa,\al)=
\frac{1}{R}I_1(\kappa)-\frac{\Delta_\alpha}{12R^2},
\\
&
\Omega_{1,3\atop 3,1}(\kappa,\al)
=
\frac{1}{R}I_3(\kappa)
-\frac{\Delta_\alpha}{6R^3}I_1(\kappa)
+\frac{\Delta_\alpha^2}{144R^4}
+\frac{c+5}{1080R^4}\Delta_\alpha
\mp \frac{\Delta_\alpha}{360R^4}d_\alpha,
\\
&\Omega_{1,5\atop 5,1}(\kappa,\al)
=
\frac{1}{R}I_5(\kappa)
-\frac{\Delta_\alpha}{4R^3}I_3(\kappa)
+\left(\frac{\Delta_\alpha^2}{48R^5}
+\frac{c+11}{360R^5}\Delta_\alpha
\right)I_1(\kappa)
\\
&\quad 
-\frac{\Delta_\alpha^3}{1728R^6}
-\frac{13(c+35)}{90720R^6}\Delta_\alpha^2
-\frac{2c^2+21c+70}{60480R^6}\Delta_\alpha
\\
&\quad \mp
\left(
\frac{\Delta_\alpha}{120R^5}I_1(\kappa)
-\frac{1}{1440R^6}
\Delta^2_\alpha-\frac{c+7}{7560R^6}\Delta_\alpha
\right)d_\alpha,
\\
&\Omega_{3,3}(\kappa,\al)
=
\frac{1}{R}I_5(\kappa)
-\frac{\Delta_\alpha}{4R^3} I_3(\kappa)
+\left(\frac{\Delta_\alpha^2}{48R^5}+\frac{c+2}{360R^5}
\Delta_\alpha+\frac{c+2}{1440R^5}\right)I_1(\kappa)
\\
&\quad
-\frac{1}{1728R^6}\Delta^3_\alpha
-\frac{5c-14}{18144R^6}\Delta^2_\alpha
-\frac{10c^2+37c+70}{362880R^6}\Delta_\alpha
-\frac{1/2c^2+c}{36288R^6}\,.
\end{align*}
Here
$I_{2n-1}(\kappa)$ are given in \eqref{I1}--\eqref{I5},
and 
\begin{align}
d_\alpha=\frac{\nu(\nu-2)}{\nu-1}(\alpha-1)
=\textstyle{\frac 1 6}
\sqrt{(25-c)(24\Delta _{\al}+1-c)}\,.
\label{Dodd}
\end{align}
These structures exhibit a remarkable 
consistency with our fermionic picture.

\section{ Final results and conclusions}\label{final}

Now we clearly see the 
structure of our fermions in the CFT limit. 
They naturally split into two parts:
\begin{align}
\betab _{2m-1}^{*}=
D_{2m-1}(\al)\betab _{2m-1}^{\mathrm{CFT}*},
\qquad \gammab _{2m-1}^{*}=
D_{2m-1}(2-\al)\gammab_{2m-1}^{\mathrm{CFT}*}\,,
\end{align}
the multipliers $D_{2m-1}(\al)$, 
$D_{2m-1}(2-\al)$ absorb all the transcendental
dependence on $\al$, the operators 
$\betab _{2m-1}^{\mathrm{CFT}*}$, 
$\gammab _{2m-1}^{\mathrm{CFT}*}$ are purely CFT-objects.

The fermions act between different Verma modules. 
In order to stay in one Verma module
it is convenient to introduce the bilinear combinations 
of fermions
\begin{align}
&\phib _{2m-1,2n-1}^{\mathrm{even}}  
=(m+n-1)
\frac{1}{2}
\left(\betab _{2m-1}^{\mathrm{CFT}*}\gammab _{2n-1}^{\mathrm{CFT}*} 
+\betab _{2n-1}^{\mathrm{CFT}*}\gammab _{2m-1}^{\mathrm{CFT}*}\right)
\nn\\
&\phib_{2m-1,2n-1}^{\mathrm{odd}}=d_{\al}^{-1}
(m+n-1)
\frac{1}{2}
\left(
\betab _{2n-1}^{\mathrm{CFT}*}
\gammab _{2m-1}^{\mathrm{CFT}*}
-
\betab_{2m-1}^{\mathrm{CFT}*}
\gammab _{2n-1}^{\mathrm{CFT}*} 
\right)
\,. 
\nn
\end{align}

The Verma module has a basis consisting of the vectors 
\begin{align}
\mathbf{i}_{2k_1-1}\cdots \mathbf{i}_{2k_p-1}\mathbf{l}_{-2l_1},\cdots  
\mathbf{l}_{-2l_q}\bigl(\phi_\al\bigr)\,.
\label{vec}
\end{align}
Conjecturally the same space 
is also created by the action of the 
$\mathbf{i}_{2k-1}$'s and fermions:
\begin{align}
\mathbf{i}_{2k_1-1}\cdots \mathbf{i}_{2k_p-1}
&\phib _{2m_1-1,2n_1-1}^{\mathrm{even}}
\cdots 
\phib _{2m_r-1,2n_r-1}^{\mathrm{even}}
\phib _{2\bar{m}_1-1,2\bar{n}_1-1}^{\mathrm{odd}}
\phib _{2\bar{m}_s-1,2\bar{n}_s-1}^{\mathrm{odd}}
\bigl(\phi _{\al}\bigr)\,.
\label{vec2}
\end{align}
For small degrees, 
the transition coefficients between \eqref{vec}
and \eqref{vec2}, modulo descendants of the $\mathbf{i}_{2k-1}$,  
can be determined 
by taking the expectation values with $\kappa=\kappa'$ 
and equating like powers of $\kappa$.  
Abbreviating $\phi_\alpha$ and writing 
$\Delta_\alpha$ as $\Delta$, we find 
\begin{align}
&\phib_{1,1}^{\mathrm{even}}\equiv
\mathbf{l}_{-2},\label{bcl}
\\
&\phib_{1,3}^{\mathrm{even}}\equiv\mathbf{l}_{-2}^2 
+\frac{2c-32}{9}\  \mathbf{l}_{-4},
\nn\\
&\phib_{1,3}^{\mathrm{odd}}\equiv
\frac{2}{3}\ \mathbf{l}_{-4},
\nn\\
&\phib_{1,5}^{\mathrm{even}}\equiv
\mathbf{l}_{-2}^3
+\frac{c+2-20\Delta+2c\Delta}{3(\Delta+2)}\ 
\mathbf{l}_{-4}\mathbf{l}_{-2}
\nn\\
&\quad +
\frac{-5600\Delta+428c\Delta-6c^2\Delta
+2352\Delta^2-300c\Delta^2+12c^2\Delta^2+896\Delta^3-32c\Delta^3}
{60\Delta(\Delta+2)}\ 
\mathbf{l}_{-6}
\,,
\nn\\
&\phib_{1,5}^{\mathrm{odd}}\equiv
\frac{2\Delta}{\Delta+2}\ \mathbf{l}_{-4}\mathbf{l}_{-2}
+\frac{56-52\Delta-2c+4c\Delta}{5(\Delta+2)}\ 
\mathbf{l}_{-6}
\,,
\nn\\
&\phib_{3,3}^{\mathrm{even}}\equiv
\mathbf{l}_{-2}^3
+\frac{6+3c-76\Delta+4c\Delta}{6(\Delta+2)}\ \mathbf{l}_{-2}\mathbf{l}_{-4}
\nn\\
&\quad +
\frac{-6544\Delta+498c\Delta-5c^2\Delta
+2152\Delta^2-314c\Delta^2+10c^2\Delta^2-448\Delta^3+16c\Delta^3}
{60\Delta(\Delta+2)}\ 
\mathbf{l}_{-6}
\,,
\nn
\end{align}

At the next degree, there are $5$ Virasoro descendants
$\mathbf{l}_{-2}^4$, $\mathbf{l}_{-4}\mathbf{l}_{-2}^2$,  $\mathbf{l}_{-4}^2$, $\mathbf{l}_{-6}\mathbf{l}_{-2}$, 
$\mathbf{l}_{-8}$, 
which are polynomials in $\Delta_{\kappa+1}$ of 
degree $4,2,1,1,0$, respectively. 
With the data at hand, 
obtained from 
the primary field $\phi_{\kappa +1}$, 
there remains one parameter undetermined. 
This can be fixed considering the first descendent 
$L_{-1}\phi _{\kappa +1}$ 
which we hope to do in future.
Nevertheless we have checked that the determinant, 
\begin{align*}
\betab _{1}^{\mathrm{CFT}*}\betab _{3}^{\mathrm{CFT}*}
\gammab _{3}^{\mathrm{CFT}*}\gammab _{1}^{\mathrm{CFT}*}\,,
\end{align*}
after subtracting a suitable multiple of $\mathbf{l}_{-2}^4$, 
has the correct degree $2$ in $\Delta_{\kappa+1}$. 
We regard it as a further supporting evidence  
in favour of the fermionic structure. 

Let us pass to conclusions.

We believe that the fermionic description 
will provide new results for
the theory of integrable models. 
For example, there is an obvious similarity 
between our
fermions and those introduced in \cite{BBS}. 
With the formulae \eqref{bcl} at hand, 
it should be possible to upgrade the qualitative description 
of form factors of descendants in \cite{BBS} 
to a quantitative level.
We hope to explain this in future works. 
Here, however, we would like to emphasise that, 
even for CFT, the fermionic description must give 
something completely new. 
Let us explain that. 

Consider the functional $Z_R^{\kappa,\kappa'}$ 
with $\kappa=\kappa'$. 
It describes the three point function for descendants of 
$\phi _\al$ and
two primary fields $\phi_{-\kappa +1}$, $\phi_{\kappa +1}$
of equal dimension $\Delta_{-\kappa+1}=\Delta_{\kappa+1}$. 
It was said several times that 
the construction generalises if we replace 
the asymptotic states described by $\phi _{\kappa +1}$,
by any other eigenstate of the
integrals of motion $I_{2n-1}$. 
The only change is that the function
$\omega$ is to be computed for the new asymptotic condition. 
It is assumed \cite{BLZI,BLZII} 
that the joint spectrum
of $I_{2n-1}$ is simple, so, in this way we 
compute all the
three-point functions for a descendant of $\phi _\al$
and descendants of $\phi_{-\kappa +1}$, $\phi_{\kappa +1}$ 
provided the latter
are eigenstates of the integrals of motion. Notice that
the descendant of $\phi _{\kappa +1}$ can be 
very deep in the Verma module. In that case the usual 
CFT computation
is rather hard to perform. 
Let us be more precise appealing to the classical limit.

In the classical limit $\nu\to 1$, the eigenstates of 
$I_{2n-1}$ 
are in correspondence with the periodic solutions of 
the KdV equation. 
Let us give some explanation about this point. 

Consider the classical KdV hierarchy with the second Poisson
structure:
$$
\{u(y_1),u(y_2)\}=
2(u(y_1)+u(y_2))\delta '(y_1-y_2)+\delta'''(y_1-y_2)\,.
$$
The integrability of the KdV equation 
is due to existence of the auxiliary linear problem:
$$
(\partial _y^2+u(y))\psi (y,\la)=\la ^2\psi (y,\la)\,.
$$
We consider the periodic case $u(y+2\pi R)=u(y)$. 
In this case
one defines the monodromy 
matrix $M(\la)$ for the auxiliary linear 
problem in a standard way. 
Then the local integrals of motion in 
involution are found in the asymptotical expansion of
$T^\mathrm{cl}(\la)=\Tr M(\al)$ for $\la ^2\to +\infty$:
$$
\log (T^\mathrm{cl}(\la))\simeq 2\pi R \la +
\sum\limits _{n=1}^{\infty} C^\mathrm{cl}I^\mathrm{cl}_{2n-1}
\la ^{-(2n-1)}\,,
$$
where
$$
C_n^\mathrm{cl}=-\sqrt{\pi}\ 
\frac {\Gamma\(\frac {2n-1} 2\)}{n!}\,.
$$
The integrals of motion $I^\mathrm{cl}_{2n-1}$ are well-known 
functionals of $u(y)$. Here we use for them the normalisation of
\cite{BLZI}. So, the classical limit is
$$
T(-iy)\ \to\  \frac 1 {1-\nu} u(y)\, 
$$
It 
brings the Virasoro commutation relations to the second
Poisson structure of the KdV hierarchy, 
and ensures the finite limits
$$
(1-\nu)^nI_{2n-1}\ \to \ 
I^\mathrm{cl}_{2n-1}\,.
$$

It is well known that periodic solutions of KdV are 
in correspondence with 
hyper-elliptic Riemann surfaces
which are two-fold covering of the Riemann sphere of $\la ^2$. 
In particular, 
the solution corresponding after the quantisation to 
the primary field
$\phi _{\kappa +1}$ corresponds to the Riemann surface of
genus 0:
$$
\mu ^2=\la ^2-\kappa ^2\,.
$$
From the point of view of classical theory, 
this is a completely trivial case 
which describes a constant solution of KdV. 
This case becomes non-trivial after the quantisation,  
because KdV is a theory with infinitely many degrees of
freedom, 
and quantising the simplest classical solution one has 
to take into account infinitely many 
zero oscillations (see \cite{QC} for a relevant discussion).
Still it is rather unpleasant to be able to quantise only 
trivial classical solutions. 
The consideration of usual, 
low-lying  descendants of $\phi_{\kappa +1}$
does not change the situation seriously: 
they describe excitations for the same classical solution. 
What are really interesting solutions in the classical case?  
They correspond 
to other Riemann surfaces. 
The simplest one is described by the elliptic curve:
$$
\mu ^2=(\la ^2-\la_1 ^2)
(\la ^2-\la _2 ^2)(\la ^2-\la _3 ^2)\,.
$$
At the quantum level,  
this solution corresponds to the following
distribution of the Bethe roots over the 
real axis in the plane of $\la ^2$. 
Going from $\la ^2=-\infty$ we first have no Bethe roots.  
Then there is a large interval 
where the Bethe roots are dense. 
Then there is a large interval without the Bethe roots, 
wherein we find one or several zeros of 
$T^\mathrm{sc}(\la,\kappa)$.
Then starting from certain point and up to 
$\la ^2 =\infty$, the Bethe roots are again dense.

For a reader who is not familiar
with periodic solutions of KdV, 
it is useful 
to think about this solution as a periodic analogue 
of one-soliton solution which we really obtain
in  the limit $R\to\infty$. 
Everybody would agree that quantising
only the trivial solutions 
when there are solitons around is
a waste of possibility.

Our fermionic construction gives a possibility to treat
this kind of asymptotic states. 
Moreover, we suppose that 
the function $\omega$ has a clear algebra-geometric 
meaning in the classical limit. 
We hope to return to all that
in one of our future publications.

\appendix

\section{General results on the asymptotics of 
$\omegab(\la,\mu|\kappa, \kappa',\al)$}

In this section we derive the asymptotic behaviour 
\eqref{asymom}
of $\omegab(\la,\mu|\kappa, \kappa',\al)$
when $\la^2,\mu^2\to\infty$. The main point 
of the argument is that in a certain domain, which we call
A-domain, the expansion \eqref{BLZ-logT} of 
$\log\rhob(\la|\kappa,\kappa')$
holds for both $\la$ and $\la q^{-1}$, and by the cancellation
due to 
\begin{align*}
\lambda^{\frac{1}{\nu}}=-(\lambda q^{-1})^{\frac{1}{\nu}}, 
\end{align*}
we have 
$\rhob(\la|\kappa,\kappa')\rhob(\la q^{-1}|\kappa,\kappa')
\simeq 1$. 
We shall suppress 
the arguments $\kappa$,  $\kappa '$ and $\al$ in
$\rhob(\la|\kappa,\kappa ')$ 
and $\omegab(\la,\mu|\kappa, \kappa',\al)$.
We set 
\begin{align*}
\rhob(\la)=
\frac {\Tb(\la,\kappa ')}{\Tb(\la,\kappa )}\,,
\qquad
\mathfrak{a}^\mathrm{sc}(\la)
=\frac{\Qb(\la q,\kappa)}
{\Qb(\la q^{-1},\kappa)}\,.
\end{align*}
For $\omegab(\la,\mu)$, 
after simple computations we get:
\begin{align}
\omegab(\la,\mu)=
&\(f_\mathrm{left}\star f_\mathrm{right}
+f_\mathrm{left}\star           
\Rdr
\star f_\mathrm{right}\)(\la,\mu)
 +
\delta ^-_\la\delta ^-_\mu
\Delta _\la^{-1} \psi(\la/\mu,\al)\,.
\label{eqomega}
\end{align}
The symbol $\star$ stands for integration over the contour 
$\gamma$ 
going clockwise around the 
zeros of $\Qb(\la,\kappa)$ with the measure
$$
dm(\theta)=\frac{d\theta ^2}
{\theta^2\rhob(\theta)(1+\mathfrak{a}^\mathrm{sc}(\theta))}\,.
$$
Here again 
$\star$, $dm(\la)$ and $\Rdr$
are slightly different than those used in 
section \ref{rhomega} or section \ref{asymp-omega}, 
but this should not cause any confusion. 

The measure $dm(\la)$ has simple poles at the zeros 
of $Q^\mathrm{sc}(\la,\kappa)$ and 
$T^\mathrm{sc}(\la,\kappa')$. 
For simplicity of presentation, 
we assume $\kappa$ and $\kappa'$ are close enough so 
that any zero of $T^\mathrm{sc}(\la,\kappa')$ 
is smaller than any zero of $Q^\mathrm{sc}(\la,\kappa)$. 

In this section if we say $f(\la)\simeq g(\la)$ 
on some half line 
of $\arg(\la ^2)$, it  means 
$f(\la)=g(\la)+O(|\la|^{-N})$ for all $N$ 
there.

From \cite{BLZII} one concludes that
\begin{align}
&\log\mathfrak{a}^\mathrm{sc}(\la)
=-F_+(\arg (\la^2))|\la| ^{\frac 1 \nu}+O(|\la| ^{-\frac 1 \nu}),\ \ \ \ 0<\arg (\la ^2) <\pi\,,
\nn\\
&\log\mathfrak{a}^\mathrm{sc}(\la)
=\ \ F_-(\arg (\la^2))|\la| ^{\frac 1 \nu}
+O(|\la| ^{-\frac 1 \nu}),\ \ \ \  -\pi<\arg (\la ^2) <0\,,
\nn
\end{align}
where $F_{\pm}(\arg(\la^2))$ 
are some functions taking 
positive values in corresponding domains. 
Hence $\mathfrak{a}^\mathrm{sc}(\la)$ decays rapidly in the upper 
half plane and grows rapidly in the lower half plane. 

Following \cite{BLZII} we write in the integral over 
the upper bank in  \eqref{eqomega} using
$$
\frac 1 {1+\mathfrak{a}^\mathrm{sc}(\eta)}
=1-\frac 1 {1+\bar{\mathfrak{a}}^\mathrm{sc}(\eta)},
\quad \bar{\mathfrak{a}}^\mathrm{sc}(\eta)
=\frac{1}{\mathfrak{a}^\mathrm{sc}(\eta)}\,,
$$
in order to separate the rapidly decreasing part. 
To formalise the story
we introduce the notation
$$
f\star g = f\circ g-f\ast g
$$
where
\begin{align}
&f\circ g=
\int\limits _{\sigma ^2}^{e^{i0}\infty} 
f(\la)g(\la)\frac {d\la ^2}{ \la ^2 \rhob (\la)}\,,
\quad f\ast g=\int\limits _{\sigma ^2}^{\infty} f(\la)g(\la)d\widetilde{m}(\la)\,.\nn
\end{align}
Here 
$\sigma ^2$ is an arbitrary point lying between 
the largest zero of $T(\al,\kappa')$ and the smallest
zero of $\Qb(\la,\kappa)$, and the modified measure is
$$
d\widetilde{m}(\la)=\frac {d\la ^2}{
\la ^2 \rhob(\la)
}
\(
\frac 1 {1+\bar{\mathfrak{a}}^\mathrm{sc}(\la e^{i0})}+
\frac 1 {1+\mathfrak{a}^\mathrm{sc}(\la e^{-i0})}\)\,.
$$

Introduce the resolvent $R$ by the equation
$$
R-K_{\al} \circ R=K_{\al}\,,
$$
and two "dressed" kernels:
\begin{align}
F_\mathrm{left}=f_\mathrm{left}+f_\mathrm{left}\circ R\,, \qquad F_\mathrm{right}=f_\mathrm{right}
+R\circ f_\mathrm{right}\,.\label{defF}
\end{align}
where the functions 
$f_\mathrm{left}(\la,\eta)$, 
$f_\mathrm{right}(\eta,\la)$ 
are defined in \eqref{flr}. 
They are singular at $\eta ^2=\la ^2$.
According to our general prescription we understand real $\la ^2$ in them 
as $\la ^2 e^{-i0}$ and then continue analytically.
The equation for the resolvent takes the form
$$
\Rdr+R\ast \Rdr
=R\ast\Rdr +\Rdr
=R\,,
$$
and the definition of $\omega$ can be rewritten as
\begin{align}
&\omegab(\la,\mu)=\omega^{(1)}(\la,\mu)+\omega^{(2)}(\la,\mu)\,,\nn\\
&\omega^{(1)}(\la,\mu)=(-F_\mathrm{left}\ast F_\mathrm{right}+
F_\mathrm{left}\ast 
\Rdr
\ast F_\mathrm{right})(\la,\mu)\,,\nn\\
&\omega^{(2)}(\la,\mu)=(f_\mathrm{left}\circ F_\mathrm{right})(\la,\mu)+
\delta ^-_\la\delta ^-_\mu\Delta _\la^{-1} \psi(\la/\mu,\al)\,.\nn
\end{align}
Now we are ready to study the asymptotical behaviour. 
We shall consider $\la ^2$ and $\mu ^2$ in the A-domain defined
as follows : 
$\pi (2\nu -1)<\arg(\la^2),\arg(\mu ^2)<\pi$. 
We prove the correct asymptotic 
behaviour there, then assume 
that it is valid for all 
$-\pi <\arg(\la^2),\arg(\mu ^2)<\pi$. 
The latter assumption is not even necessary for our goals,
but we do not see why it should not be true having 
in mind that the only infinite series of 
poles of $\omega(\la,\mu)$ 
are the zeros of $\Tb(\la,\kappa)$
which accumulate to $\la ^2=-\infty$.

The importance of A-domain is due to the fact that in it
\begin{align}
\rho (\la)\rho (\la q^{-1})\simeq 1\,.\label{rhorho}
\end{align}
Introduce the operation
\begin{align}
\delta ^+_\la f(\la)=f(\la )+\rho (\la)f(\la q^{-1})\,.
\label{deltap}
\end{align}
Using the definitions it is not hard to show that for $\la ^2$, $\mu ^2$ in A-domain
\begin{align}
\delta ^+_\la F_\mathrm{left}(\la,\eta)\simeq 0,\quad \delta ^+_\mu F_\mathrm{right}(\eta,\mu)\simeq 0\,.\nn
\end{align}
These equations imply
\begin{align}
&F_\mathrm{left}(\la,\eta)\simeq\sqrt{\rho (\la)}\sum\limits _{k=1}^{\infty}\la ^{-\frac {2k-1}{\nu}}
F_{\mathrm{left},\ k}(\eta)\label{asymF}\,,\\
&F_\mathrm{right}(\eta, \mu)\simeq\sqrt{\rho (\mu)}\sum\limits _{k=1}^{\infty}\mu ^{-\frac {2k-1}{\nu}}
F_{\mathrm{right},\ k}(\eta)\nn\,.
\end{align}
It is easy to argue that $F_{\mathrm{left},\ k}(\eta)$, $F_{\mathrm{right},\ k}(\eta)$ grow for $\eta \to\infty$
as powers of $\eta$. This is enough to ensure that the "connected part" $\omega^{(1)}(\la,\mu)$
has the desired asymptotics: we just substitute the asymptotics \eqref{asymF} into the formula for
$\omega^{(1)}(\la,\mu)$ and observe that all the integrals converge because of exponential in $\eta$
decay of $d\widetilde{m}(\eta)$. With 
the "disconnected part" $\omega^{(2)}(\la,\mu)$ the situation is far more delicate, studying it
we shall understand the importance of the term 
$\delta^-_\la\delta^-_\mu\Delta_\la^{-1}\psi(\la/\mu,\al)$ 
in 
$\omega(\la,\mu)$.

Let us  evaluate the last term in $\omega^{(2)}(\la,\mu)$
considering $\la^2,\mu^2>\sigma^2$. Using the definition  \eqref{Dinv} it is easy to see that
\begin{align}
&\delta^{-}_\mu \Delta_{\lambda}^{-1}\psi(\la/\mu)
=-\int\limits _{0}^{e^{i0}\infty }
\frac 1 {2\nu \bigl(1+(\la/\eta)^{\frac 1 \nu}\bigr)}
f_\mathrm{right}(\eta,\mu)
\ \frac {d\eta ^2}{2\pi i \eta^2}+
\delta ^+_\mu
\frac 1 {4\nu \bigl(1-(\la/\mu)^{\frac 1 \nu}\bigr)}\,.\nn
\end{align}
The function $F_\mathrm{right}(\eta,\mu)$ allows analytical
continuation with respect to $\eta$, so, we shall use it for
all $\eta^2\in \mathbb{R}_+$. 
Substite
$f_\mathrm{right}=F_\mathrm{right}-K_\al\circ F_\mathrm{right}$ 
and compute the integral
\begin{align}
&\int\limits _
{0}^{\infty }
\frac {K_\al (\theta,\eta)}{2\nu(1+(\la/\theta)^{\frac 1\nu})}
\ \frac {d\theta ^2}{ \theta ^2}
=-\psi (\la/\eta,\al)
-\frac 1{2\nu(1-(\la/\eta)^{\frac 1 \nu})}\,.
\nn
\end{align}
We get after some straightforward computations
\begin{align}
\omega^{(2)}(\la,\mu)=\omega^{(3)}(\la,\mu)+\omega^{(4)}(\la,\mu)\,,\nn
\end{align}
where
\begin{align}
&\omega^{(3)}(\la,\mu)
=-\int\limits _0^{\sigma ^2}\delta ^-_\la 
\frac 1 {2\nu \bigl(1+(\la/\eta)^{\frac 1 \nu}\bigr)}
\cdot F_\mathrm{right}(\eta,\mu)\ \frac {d\eta ^2}{2\pi i \eta^2}\,,\label{eqW1}
\\
&\omega^{(4)}(\la,\mu)=-VP\int\limits _{\sigma ^2}^{\infty}
\delta ^-_\la\delta ^+_\eta 
\frac 1 {2\nu \bigl(1-(\la/\eta)^{\frac 1 \nu}
\bigr)}\cdot F_\mathrm{right}(\eta,\mu)
\frac {d\eta ^2}{2\pi i \eta^2\rho (\eta)}\,,\label{eqW}
\end{align}
where 
the principal value refers 
to the pole at $\eta ^2=\mu ^2$. 

For $\omega^{(3)}(\la,\mu)$ the asymptotics 
of the kind \eqref{asymom}
follows immediately from \eqref{asymF} and
$$ \delta ^+_\la\delta ^-_\la 
\frac 1 {2\nu \bigl(1+(\la/\eta)^{\frac 1 \nu}\bigr)}\simeq 0\,. $$

Consider $\omega^{(4)}(\la,\mu)$.
We check the equations
\begin{align}
\delta ^+_\la \omega^{(4)}
(\la,\mu)\simeq 0,\qquad \delta ^+_\mu 
\omega^{(4)}(\la,\mu)\simeq 0\,.
\label{dW}
\end{align}
The first of them follows immediately 
from two facts. 
First, 
$$
\delta ^+_\la\delta ^-_\la\delta ^+_\eta \frac 1 {2\nu \bigl(1-(\la/\eta)^{\frac 1 \nu}\bigr)}\simeq 0\,.
$$
Second, writing explicitly 
\begin{align}
\delta ^-_\la\delta ^+_\eta \frac 1 {2\nu \bigl(1-(\la/\eta)^{\frac 1 \nu}\bigr)}=
\frac {\rho(\eta)-\rho(\la)}{2\nu \bigl(1-(\la/\eta)^{\frac 1 \nu} \bigr)}
+\frac {1-\rho(\la)\rho(\eta)}{2\nu \bigl(1+(\la/\eta)^{\frac 1 \nu} 
\bigr)}\,,
\label{ggg}
\end{align}
and recalling the asymptotic expansion for $\rho (\la)$,
$\rho (\eta)$,  
we see that
asymptotically 
for $\la ^2\to +\infty$, $\eta ^2\to +\infty$ 
the singularities in \eqref{ggg}
disappear. 
Altogether we have for the asymptotics in both arguments
\begin{align}
\delta ^-_\la\delta ^+_\eta \frac 1 {2\nu \bigl(1-(\la/\eta)^{\frac 1 \nu}\bigr)}\simeq \sqrt{\rho (\la)}
\sum _{m,n=1}^{\infty}\la ^{-\frac {2m-1} \nu}\eta ^{-\frac n \nu}C_{m,n}
\label{asymreg}\,.
\end{align}
To prove the second equation 
in \eqref{dW} it is not sufficient to use 
\eqref{asymF}  because $F_\mathrm{right}(\eta,\mu)$
has simple poles at 
$\eta ^2=\mu ^2$ and $\eta ^2=\mu ^2 q^2$ 
which contribute to
the analytic continuation $\mu\to\mu q^{-1}$. 
However, it is easy to
see that the corresponding contributions cancel. 

Using the first of equations \eqref{dW}
we get 
$$
\omega^{(4)}(\la,\mu)\simeq \sqrt{\rho (\la)}\sum\limits _{m=1}^{\infty}\la ^{-\frac {2m-1}\nu}
\omega^{(4)}_m(\mu)\,,
$$
where due to \eqref{asymreg} the functions $\omega^{(4)}_m(\mu) $ 
are given
by convergent integrals. 
These functions satisfy 
$\delta ^+_\mu \omega^{(4)}_m(\mu)\simeq 0$, 
and  do not grow for $\mu ^2\to+\infty$.  
Hence  $\omega^{(4)}(\la,\mu)$ has the asymptotics 
of the kind \eqref{asymom}.

\bigskip

\noindent

{\it Acknowledgements.}\quad
HB is grateful to 
the Volkswagen Foundation for financiall support.
Research of MJ is supported by the Grant-in-Aid for Scientific 
Research B-20340027. 
Research of TM is supported by the Grant-in-Aid for Scientific Research
B-17340038.
Research of FS is supported by  RFBR-CNRS grant 09-02-93106 and
by EU-grant MEXT-CT-2006-042695 during his visit to DESY, Hamburg.

FS thanks Masaki Kashiwara for inviting him to
RIMS where the most important part of this research was carried out.
This visit was supported by the Grant-in-Aid for Scientific Research B-18340007.

HB would like to thank F. G\"ohmann, A. Kl\"umper 
and K. Nirov for the stimulating discussions. 
FS would like to thank S. Lukyanov for many valuable discussions.

\end{document}